\documentclass[twocolumn,showpacs,prb,superscriptaddress,aps,floatfix]{revtex4-1}

\usepackage{rotating}
\usepackage{amssymb,amsmath,mathtools}
\usepackage{tabularx}
\usepackage{color}
\usepackage{dsfont}
\usepackage{graphicx}
\usepackage[latin9]{inputenc}
\usepackage{floatrow}
\usepackage{array,multirow}

\usepackage[active]{srcltx}
\usepackage[sort&compress]{natbib}
\usepackage[dvips]{epsfig}
\usepackage{float}
\usepackage[normalem]{ulem}
\usepackage[dvipsnames]{xcolor}
\usepackage[colorlinks,linkcolor=blue,citecolor=blue,urlcolor=blue]{hyperref}

\newcommand{\val} {Institute of Materials Science (ICMUV), University of Valencia, Catedr\'{a}tico Beltr\'{a}n 2, E-46980, Valencia, Spain}
\newcommand{\cnrism} {Istituto di Struttura della Materia and Division of Ultrafast Processes in Materials (FLASHit) of the National Research Council, via Salaria Km 29.3, I-00016 Monterotondo Stazione, Italy}
\newcommand{\etsf} {European Theoretical Spectroscopy Facilities (ETSF)}
\newcommand{\tov} {Dipartimento di Fisica, Universit\`{a} di Roma Tor Vergata, Via della Ricerca Scientifica 1, 00133 Rome, Italy}
\newcommand{\DISC} {DISC, Dipartimento di Scienze Chimiche, University of Padova,Via Marzolo 1, I-35131 Padova, Italy}
\newfloatcommand{capbtabbox}{table}[][\FBwidth]

\renewcommand{\[}{\left[}
\renewcommand{\]}{\right]}
\renewcommand{\(}{\left(}
\renewcommand{\)}{\right)}

\def\la         {\langle}
\def\ra         {\rangle}

\def\wl         {\Big\{}
\def\wr         {\Big\}}
\newcommand{\Langle}{\left\langle}
\newcommand{\Rangle}{\right\rangle}
\def\dn         {\downarrow}
\def\up         {\uparrow}
\def\rar        {\rightarrow}  
\def\Rar       {\Longrightarrow}

\newcommand{\olrar}[1]{\overleftrightarrow{#1}}
\newcommand{\orar}[1]{\overrightarrow{#1}}
\newcommand{\olar}[1]{\overleftarrow{#1}}
\newcommand{\eq}[1]{\begin{align}#1\end{align}}
\newcommand{\ml}[1]{\begin{multline}#1\end{multline}}
\newcommand{\eqg}[1]{\begin{gather}#1\end{gather}}
\newcommand{\seq}[1]{\begin{subequations}#1\end{subequations}}
\newcommand{\wh}[1]{\widehat{#1}}

\newcommand{\bs}[1]{\mathbf{#1}}
\newcommand{\uu}[1]{\underline{#1}}
\newcommand{\oo}[1]{\overline{#1}}
\newcommand{\e}[1]{Eq.\eqref{#1}}

\newcommand{\h}[1]{\hat{#1}}

\newcommand{\wt}[1]{\widetilde{#1}}

\def\nt {\bar{n}}
\def\mt {\bar{m}}

\def\ga         {\alpha}

\def\gd         {\delta}

\def\gee        {\epsilon}
\def\gl         {\lambda}

\def\go         {\omega}

\def\gr         {\rho}
\def\gs         {\sigma}

\def\AA		{{\bf A}}

\def\GG		{{\bf G}}

\def\pp		{{\bf p}}
\def\rr		{{\bf r}}

\def\II		{{\bf I}}
\def\JJ		{{\bf J}}
\def\LL		{{\bf L}}
\def\MM		{{\bf M}}

\def\kk		{{\bf k}}
\def\qq		{{\bf q}}

\newcommand{\w}{\omega}
\renewcommand{\ss}{\mathbf{s}}
\newcommand{\xx}{\mathbf{x}}

\newcommand{\bea}{\begin{eqnarray}}
\newcommand{\eea}{\end{eqnarray}}

\begin{document}

\title{Spinorial formulation of the GW-BSE equations and spin properties of excitons in 2D Transition Metal Dichalcogenides}

\author{Margherita Marsili}
\affiliation{\DISC} 

\author{Alejandro Molina-S\'{a}nchez}
\affiliation{\val}

\author{Maurizia Palummo}
\affiliation{\tov} 
\affiliation{\etsf} 

\author{Davide Sangalli}
\affiliation{\cnrism} 
\affiliation{\etsf} 

\author{Andrea Marini}
\affiliation{\cnrism} 
\affiliation{\etsf} 

\date{\today}
\begin{abstract}
In many paradigmatic materials, like Transition Metal Dichalcogenides, the role played by the
spin degrees of freedom is as important as the one played by the electron-electron interaction. 
Thus an accurate treatment of the two effects and of their interaction is necessary for an accurate and predictive study of the optical and electronic properties of these materials.
Despite the GW-BSE approach correctly accounts for electronic correlations the
spin--orbit coupling effect is often neglected or treated perturbatively. 
Recently spinorial formulations of GW--BSE have become available in different flavours in
material--science codes. 
Still an accurate validation and comparison of different approaches is missing.
In this work we go through the derivation of non collinear GW--BSE approach. 
The scheme is applied to transition metal dichalcogenides comparing perturbative and full spinorial approach.
Our calculations reveal that dark--bright exciton splittings are generally improved when the spin orbit coupling 
is included non perturbatively. 
The exchange-driven intravalley mixing between the A and B exciton is found
to be extremely important in the case of MoSe$_2$. 
We finally define the excitonic spin and use it 
to sharply analyze the spinorial properties of Transition Metal
Dichalcogenides excitonic states.
\end{abstract}
\pacs{ PACS }
\maketitle

\section{Introduction}\label{sec:intro}
The investigation of the excited state properties of materials by means of modern {\it Ab--Initio} theories is a rapid developing field, that yielded 
notable progresses in our understanding of bulk, surfaces, nanostructures,  molecules and disordered systems~\cite{Onida2002}. 
At the same time an increasing number of experimental and technological applications are constantly pointing out 
the central role played by the spin degrees of freedom in the explanation of 
novel and intriguing physical processes. An example is the chirality effect, observed in
low--dimensional systems lacking structural inversion symmetry~\cite{Bode2007/05/10/online}, or
the fact that spin-orbit coupling (SOC) in several cases, like for the class of Transition Metal Dichalcogenides (TMD) or Topological Insulators (TI) 
but also 
nano-materials with light atoms, like Carbon nanotubes,
results to be essential to correctly describe their electronic and optical excitations\cite{Kuemmeth2008/03/27/online}.
Moreover an accurate description of the spin dynamics is essential in the fields of
spintronics and magneto--optics~\cite{Picozzi2006}.
Despite this fact,
most of the many-body calculations based on the GW and Bethe-Salpeter-Equation (BSE) methods \cite{Onida2002}  
have been carried out within a spin--indipendent or spin--polarized~\cite{rodl_prb2008} framework, generally neglecting
the SOC. 
Presently several electronic structure codes \cite{Sangalli_2019,Wu2019,Nielsen_2020,vasp}, that include SOC within the GW-BSE approach 
either perturbatively \cite{Deslippe2012,Qiu_prl13} 
or fully considering the spinorial nature of the electronic wave functions, have been applied in a number of cases
\cite{Molina-Sanchez2016, Molina-S'anchez2013,Giorgi2018,Palummo2015_NL,Guilhon_PhysRevB_2019,Wu2019,Nielsen_2020,Deilmann_PRL2020}.
However the mathematical derivation and a complete discussion of the spinorial formulation of GW--BSE is not present in the literature and
moreover an accurate comparison of the full spinorial GW--BSE with the perturbative approach, benchmarking the latter, is also missing.

Besides their importance 
for fundamental and applicative reasons \cite{Wang_RevModPhys2018,koperski_nanophot2017,Manzeli2017}, transition metal dichalcogenide (TMD) 
monolayers are ideal for testing the different level of SOC inclusion
because the presence of strong spin-orbit interaction
and the enhancement of many-body effects are at the basis of their intriguing electronic and optical properties.
The strong spin-orbit interaction determines macroscopic
features of the optical properties of these materials, like the presence
of spin-splitted peaks and valley-selective optical transitions in
their absorption spectra. Moreover SOC is responsible of finer details
which are nevertheless crucial for applications of TMDs in opto-electronic devices. These
are the splitting and the energetic order of spin-allowed (bright)
and spin-forbidden (dark) excitons which are involved in the exciton dynamics.
In this view, TMDs is an optimal class of materials against which
testing and comparing a full and a perturbative SOC formulation
up to the GW-BSE level, by using the same set of computational parameters.
Despite most of the equations can be obtained as a direct extension of the standard formulation~\cite{Aryasetiawan2008}, we give a detailed deviation non collinear GW-BSE equations. In doing so we aim at providing a complete reference with all spin-indexes carefully included and with an extensive discussion of the different approximation chosen.

We show how this formulation allows including SOC in a natural way
at the level of the ground-state calculation and it is nicely suited for the study of the
optical properties of any many-body quantum-mechanical system where the dependence from the spin can
be described as a non-local term in the hamiltonian.
Finally we apply this formalism to the calculation of the electronic and optical properties of 
Group-VI TMD monolayers (MX$_{2}$,
with M$=$Mo, W and X$=$ S, Se) and accurately compare the no-, perturbative and full SOC schemes. 
Furthermore a full analysis of the spin character
of its excitons, relevant for envisaged valleytronics applications,
intrinsically not achievable in a perturbative approach, is presented.

\section{The non--collinear Many Body problem}\label{sec:MBSOC}

We start from the many--body total Hamiltonian of the system including first order relativistic corrections
\eq{
\wh{H}=\wh{H}^0+\wh{H}^{\(e-e\)}+\wh{H}^{\(RK\)}+\wh{H}^{\(SOI\)}.
\label{eq:MB_H}
}
Here ${\wh{H}^0}$ is the non relativistic one body hamiltonian, composed by a kinetic term and the atomic scalar external potential,
$\wh{H}^{\(e-e\)}$ the electron--electron Coulomb interaction, while ${\wh{H}^{\(RK\)}+\wh{H}^{\(SOI\)}}$ are the first order relativistic corrections.
${\wh{H}^{\(RK\)}}$ is the mass--velocity term plus the Darwin term, while ${\wh{H}^{\(SOI\)}}$ is the   Spin--Orbit--Interaction (SOI)
term~\footnote{we do not consider here external vector potentials}.

In the position, momentum and spin of each electron $\h{\xx},\h{\pp},\h{\ss}$,
the different terms can be expressed as
\seq{
\eqg{
\wh{H}^0         = H^0\[\{\h{\xx}_n,\h{\pp}_n\}\] = \sum_i h^0\(\h{\xx}_i,\h{\pp}_i\) \label{eq:MB_H0}, \\
\wh{H}^{\(e-e\)} = H^{\(e-e\)}\[\{\h{\xx}_n\}\]   = \frac{1}{2} \sum_{i\neq j} \frac{1}{|\h{\xx}_i-\h{\xx}_j|}, \\
\wh{H}^{\(RK\)}  = H^{\(RK\)}\[\{\h{\pp}_n\}\]    = \sum_i h^{\(RK\)}\(\h{\pp}_i\), \\
\wh{H}^{\(SOI\)} = \sum_i H^{\(SOI\)}\[\{\h{\xx}_n,\h{\pp}_n\}\]\(\h{\ss}_i\).  
}
\label{eq:MB_SOI_s}
}
Here bold symbols indicate spatial vectors, ${\h{h}^0} = \hat{t} + {\hat{v}^{ext}}$ is the sum of the kinetic and one-body external potential, while
${\wh{H}^{\(RK\)}}$ is the sum of one body terms and for an explicit expression of this term  we refer the reader for example to Ref.~\onlinecite{schwabl2005advanced}, eqs.~9.2.2.
By following Ref~.\onlinecite{VanYperenDeDeyne2012} we know that the SOI term is composed of three different contributions:
\eq{
\wh{H}^{\(SOI\)}=\wh{H}^{\(SO\(N\)\)}+\wh{H}^{\(SO\(2e\)\)}+\wh{H}^{\(SOO\)},
\label{eq:MB_SOI}
}
Here $\wh{H}^{\(SO\(2e\)\)}$ and $\wh{H}^{\(SOO\)}$ are many-body terms, called ``two--electrons spin--orbit'' and ``spin--other--orbit'' respectively.
${\wh{H}^{\(SO\(N\)\)}}$, or ``one--electron spin--orbit'', is a purely one body term:
\begin{eqnarray}
\wh{H}^{\(SO\(N\)\)} = \sum_i \uu{v}^{\(SOC\)}\(\h{\xx}_i,\h{\pp}_i\),  \label{eq:1B_SOC}
\end{eqnarray}
with $\uu{v}$ a $2\times 2$ matrix in the $\up,\dn$ spin space.
What is relevant here is that $H^{\(SOI\)}$ can be expressed as a sum of terms which depend
on a single spin operator and can thus be conveniently written as a sum of $2\times 2$ matrix in the $\up,\dn$  spin space. Since 
all other terms are spin--independent this implies that the whole Hamiltonian can be expressed as a sum of $2\times 2$ matrices.
We denote as $\uu{o}$ such matrices in the $\up,\dn$ spin space.

\subsection{The single--particle part of the Hamiltonian within Density Functional Theory}\label{sec:DFTSOC}
Within DFT the whole  \e{eq:MB_H} is replaced with a mean--field representation:
\eq{
 \wh{H}\Rar\sum_i \uu{h}^{KS}\[\uu{\gr}\]\(\h{\xx}_i,\h{\pp}_i\).
 \label{eq:DFT_transform} 
}
The one--body KS Hamiltonian depends on the electronic density matrix, $\uu{\gr}$ defined in \e{eq:2.5b}.
and reads
\ml{
h^{KS}_{ss'}\[\uu{\gr}\]\(\h{\xx},\h{\pp}\)=h\(\h{\xx},\h{\pp}\) \gd_{ss'}+v^{\(SOC\)}_{ss'}\(\h{\xx}\)+\\+v^{\(Hxc\)}_{ss'}\[\uu{\gr}\]\(\h{\xx}\),
\label{eq:1B_H}
}
with $s$ the spin index. In \e{eq:1B_H} $\h{h}=\h{h}^0+\h{h}^{(RK)}$ while $\h{v}^{\(SOC\)}$ is the 
one--electron spin--orbit-coupling nucleus term which comes from the $SO\(N\)$ term defined in \e{eq:1B_SOC} and $v^{\(Hxc\)}_{ss'}$ is the sum of the exchange-correlation and Hartree potential.
The more general form of $\h{v}^{\(SOC\)}$ (in case the vector potential is zero) is
\begin{align}
\uu{v}^{\(SOC\)}\(\h{\xx},\h{\pp}\)=\frac{1}{2c^2}\, \uu{\pmb{\gs}}\cdot[\pmb{\nabla} v^{\(ext\)}\(\h{\xx}\)\times \h{\pp}] ,
\label{eq:1B_H_SOC} 
\end{align}
Here $\uu{\pmb{\gs}}$ is the three dimensional vector whose components are the Pauli matrices.
In a plane wave representation the $SOC$ is, in practice, accounted for by the use of pseudo--potentials~\cite{Corso2005}, and
also $\h{h}^{(RK)}$ corrections are taken into account for the kinetic energy of core electrons in the generation
of the pseudo--potential~\footnote{Relativistic pseudo--potentials are often created directly using
the Dirac equation which includes relativistic corrections to all orders}. The SOC contribution from the pseudo--potential
captures what is called ``local SOC'', which is due to the electrons orbiting around the nuclei, it neglects instead 
the itinerant SOC which cannot be easily captured in periodic boundary conditions. Relativistic corrections to the
kinetic energy of valence (and conduction) electrons are instead usually neglected.

Neglecting the relativistic corrections $\h{v}^{\(Hxc\)}$ is the mean field replacement of $\wh{H}^{(e-e)}$.
When relativistic corrections are taken into account instead, one should in principle account for the effects
of the many--body terms $\wh{H}^{\(SO\(2e\)\)}$ and $\wh{H}^{\(SOO\)}$.
A shorcut~\cite{VanYperenDeDeyne2012} is to add a term constructed replacing $v^{ext}$
with $v^{\(Hxc\)}$ into Eq.~\eqref{eq:1B_H_SOC}.
Doing so partially accounts for the physics of $\wh{H}^{\(SO\(2e\)\)}$
while the effects described by $\wh{H}^{\(SOO\)}$ are neglected, as commonly done
in standard DFT calculations.

Thus there are two terms entering the $\h{h}^{KS}$ hamiltonian which are non diagonal in spin--space
and are responsible for inducing spinorial eigen--states: $\h{v}^{\(SOC\)}$
and $\h{v}^{\(Hxc\)}$.~\footnote{Note that one could still define a collinear functional of the sole density.~\cite{R.M.Dreizler1990}
However it is more convenient to work within non-collinear DFT which gives, not only the exact density but also the exact spin-magnetization}.
The spin-dependent formulation of the Density Functional Theory \cite{Barth1972}
in its Local Spin Density Approximation\,(LSDA) is nowdays implemented in several ab-initio codes and is at the basis of the
present excited state calculations. It gives rise to a renormalization of the SOC splitting.
In app.~\ref{App:DFT_and_SOC} we also show how the non collinear form of the hamiltonian can be rewritten
in terms of density and magnetization coupling with the xc scalar potential $\phi_{xc}$ plus an xc magnetic field $\mathbf{B}_{xc}$.
If the local magnetization is zero everywhere, then $\mathbf{B}_{xc}=0$.

The eigenstates of $\h{h}^{KS}$ are vectors in the $\up,\dn$ spin space:
\begin{align}
	\la \xx |\h{c}^{\dagger}_{\II}|0 \ra=  \overrightarrow{\psi}_\II\(\xx\)=\la \xx\orar{ | n \kk \ra}\equiv
\(
  \begin{array}{c}
    \phi_{n\kk \up}\(\xx\) \\ \phi_{n\kk \dn}\(\xx\)
  \end{array}  
\),
\label{eq:spinor_def}
\end{align}
with $n$ the spinor band index, $\kk$ a Brillouin Zone generic point, which for now on we group in the
index $\II\equiv\(n,\kk\)$ to simplify the notation.
$\overrightarrow{\psi}$ satisfies the matrix equation
\begin{align}
\uu{h}^{KS}\(\h{\xx}\)\overrightarrow{\psi}_\II\(\xx\)= \gee_\II \overrightarrow{\psi}_\II\(\xx\).
\label{eq:2.3}
\end{align}

Thanks to \e{eq:spinor_def} we can define the fermionic field operators
\begin{align}
 \wh{\Psi}\(\xx,s\)=\sum_{\II} \phi_{\II s}\(\xx\) \h{c}_{\II},
\label{eq:2.5}
\end{align}
with the operators written in the Heisenberg representation. In \e{eq:2.5} we have embodied in the spinorial wavefunction, $\phi_{\II s}$, the 
$\frac{1}{\sqrt{N_{\kk}}}$ prefactor, with $N_{\kk}$ the number of $\kk$--points. In this way we can write, for example, the spin components of the density
matrix as
\begin{align}
 \gr_{ss'}\(\xx\)=  \sum_{\II} \sum_{\ga=0,3} \phi^*_{\II s}\(\xx\) \[\gs_{\ga}\]_{ss'} \phi_{\II s'}\(\xx\),
\label{eq:2.5b}
\end{align}
where $\phi^*_{\II s}$ is the complex conjugate of $\phi_{\II s}$.
The properties of the spinorial field operators can be easily obtained from some anticommutation rules of the fermionic creation and 
annihilation operators. We have that
\begin{subequations}
  \label{eq:2.6}
\begin{gather}
 \{\h{c}_{\II_1},\h{c}^{\dagger}_{\II_2}\}=\gd_{\II_1,\II_2},\\
 \sum_s \int d\xx\, \phi^*_{\II_1 s}\(\xx\) \phi_{\II_2 s}\(\xx\)=\gd_{\II_1,\II_2},\\
 \sum_{\II} \phi^*_{\II s_1}\(\xx_1\) \phi_{\II s_2}\(\xx_2\)=\gd_{s_1s_2}\gd\(\xx_1-\xx_2\),
\end{gather}
\end{subequations}
where $\gd_{\II,\JJ}\equiv \gd_{\kk,\kk'} \gd_{n,n'}$. 

\subsection{The interaction term and the perturbative expansion}\label{sec:PT}
A crucial point of the present formulation is that by replacing the bare single--particle hamiltonian, $\uu{\h{h}}$ with the
Kohn--Sham\,(KS) hamiltonian, $\uu{\h{h}}^{KS}$ the whole perturbative expansion  is done on top of the KS energies and eigenvectors.

A subtle but essential point is that, however, in order to prevent double counting problems the 
pure electron--electron interaction  needs to be ammended.
Without relativistic corrections this means that
\begin{align}
\sum_i \h{h}^0_i +\wh{H}^{\(e-e\)}\rar \sum_i \h{h}^{KS}_i+\Delta\wh{H}^{\(e-e\)},
\label{eq:dft.1}
\end{align}
with 
\eq{
\label{eq:dft.2}
\Delta \wh{H}^{\(e-e\)}= \wh{H}^{\(e-e\)}-\sum_i\[\h{v}^{Hxc}_i\].
}
which is done in practice by subtracting ${v^{(xc)}}$ to the many body self energy $\Sigma$.

In presence of Relativistic corrections one needs to replace in \e{eq:dft.1}

\begin{eqnarray}
\h{h}^0        &\rar& \h{h}^0+\h{h}^{(RK)}+\h{v}^{SOC} \\
\wh{H}^{(e-e)} &\rar& \wh{H}^{(e-e)}+\wh{H}^{\(SO\(2e\)\)}+\wh{H}^{\(SOO\)}.
\end{eqnarray}
The resulting effective electron--electron interaction is non diagonal in spin--space and one should in principle
follow the derivation of Ref.~\onlinecite{Aryasetiawan2008} to define the proper many--body self--energy
at the diagrammatic level.
Here we neglect such complication and we focus on the effect of using a fully non--collinear non--interacting Hamiltonian,
while keeping a standard spin independent interaction at the diagrammatic level. This means that we will use the standard definition
of the many--body self--energy and still rely on \e{eq:dft.2} for applying corrections of MBPT on top of DFT.

\section{Spinorial formulation of the Hedin's equations}\label{sec:Hedin_equations_spinors}
When a non--collinear potential is present in $\uu{h}$, this implies that the whole formulation of the Many--Body problem must be rewritten in the spinorial basis.
We start from the standard Hedin equation in the space and spin, $\(\xx,s\)$, basis.
Then, using the definition of the spinorial field operators of \e{eq:2.3} we expand all terms in the theory in the spinorial basis.
In app.~\ref{App:Hedin_textbook} we give a short review of the Hedin's equations, which solve exactly the problem.
In order to rewrite the MBPT in the spinorial representation we note that, in principle
by using \e{eq:2.5}, the different components of Hedin's equations can be conveninently rotated.

In practice we define two {\em maps}, $M_2$ and $M_4$:
\ml{
M_2:F\(1,2\) \equiv \phi^*_{\II_1 s_1}\(\xx_1\) F\(1,2\) \phi_{\II_2 s_2}\(\xx_2\)\\
             = F_{\II_1\II_2}\(t_1,t_2\),\label{eq:5.0} 
}
and
\ml{
 M_4:F\(1,2,3,4\) \equiv \\\phi^*_{\II_1 s_1}\(\xx_1\)\phi_{\II_2 s_2}\(\xx_2\) F\(1,2,3,4\) \\ \phi_{\II_3 s_3}\(\xx_3\) \phi^*_{\II_4 s_4}\(\xx_4\) =
    F_{ \substack{ \II_1 \II_2 \\ \II_3 \II_4} }\(t_1,t_2,t_3,t_4\) .
\label{eq:m_4}
}
In \e{eq:5.0} and \e{eq:m_4} $n\equiv\(\xx_n,s_n,t_n\)$ and repeated variables are either integrated or summed up.
Thanks to these two maps we can easily define the representations in the spinorial basis of the different components of Hedin's equations.
More in detail we construct:
\seq{
\begin{eqnarray}
 \label{eq:map_G}
 G_{\II_1\II_2}\(t_1,t_2\)&=&M_2:G\(1,2\) ,\\
 \label{eq:map_Sigma}
 \Sigma_{\II_1\II_2}\(t_1,t_2\)&=&M_2:\Sigma\(1,2\) ,\\
 \label{eq:map_v}
 V_{  \substack{ \II_1 \II_2 \\ \II_3 \II_4} }&=&M_4:v\(1,3\)\gd\(1,2\)\gd\(3,4\) ,\\
 \label{eq:map_W}
 W_{  \substack{ \II_1 \II_2 \\ \II_3 \II_4} }\(t_1,t_2\)&=&M_4:W\(1,2\)\gd\(1,3\)\gd\(2,4\) ,\\
 \label{eq:map_gamma}
 \wt{\Gamma}_{ \substack{ \II_1 \II_2 \\ \II_3 \II_4}}\(t_1,t_2;t_3\)&=&M_4:\wt{\Gamma}\(1,2;3\)\gd\(3,4\) ,\\
 \label{eq:map_chi}
 \wt{\chi}_{  \substack{ \II_1 \II_{2} \\ \II_3 \II_{4}} } \(t_1,t_2\)&=&M_4:\wt{\chi}\(1,2\)\gd\(1,3\)\gd\(2,4\).
\end{eqnarray}
}
The kind of map to be applied depends on the number of field operators involved in the definition of the corresponding quantity and not on the number of space-time or spin arguments.
This implies the need of delta functions which extend ``contracted quantities''. For example the response function $\chi(1,2)$ is a contraction of the more
general two particles Green function $L(1,3;2,4)$ with $\wt{\chi}(1,2)=\wt{L}(1,1;2,2)$.
The proof of each rotation is given in the  Appendix\,\ref{app:rotation} and referenced here when necessary.
As a simple example we see that the most elemental ingredient of MBPT is the GF. This can be rotated by simply using \e{eq:2.5}
\begin{align}
 G\(1,2\)=\sum_{\II_1,\II_2} \phi_{\II_1 s_1}\(\xx_1\) G_{\II_1\II_2}\(t_1,t_2\)  \phi^*_{\II_2 s_2}\(\xx_2\).
 \label{eq:map_G_explicit}
\end{align}

Then the Dyson equation reads
\begin{multline}
G_{\II_1 \II_2}\(t_1,t_2\)=G^{\(0\)}_{\II_1 \II_2}\(t_1,t_2\)+G^{\(0\)}_{\II_1 \II_3}\(t_1,t_3\)\\ \times \Sigma^{Hxc}_{\II_3 \II_4}\(t_3,t_4\) G_{\II_4
\II_2}\(t_4,t_2\)
\label{eq:map_Dyson_explicit}
\end{multline}
with $\Sigma^{Hxc}=v^H+\Sigma$; the hartree potential and the self--energy are defined as
\seq{
\begin{eqnarray}
v^{H}_{\II_3 \II_4}\(t_3\) &=& -i V_{  \substack{ \II_3 \II_4 \\ \II_5 \II_{5'}} }  G_{\II_5 \II_{5'}}\(t_3,t_3^+\),
\label{eq:map_vH_explicit} \\
\Sigma_{\II_1 \II_2}\(t_1,t_2\) &=&  -i G_{\II_{1'} \II_3}\(t_1,t_3\)\times \nonumber \\
 && \wt{\Gamma}_{  \substack{ \II_3 \II_2 \\ \II_4 \II_{4'}} }\(t_3,t_2;t_4\) 
W_{  \substack{ \II_{4'} \II_{4} \\ \II_1 \II_{1'}}}\(t_4,t_1\).
\label{eq:map_Sigma_explicit}
\end{eqnarray}
}
The equation of motion for the vertex can be derived by using a generalized chain rule written in the spinorial basis, this is 
derived  in Appendix\,\ref{app:vertex}:
\begin{multline}
\wt{\Gamma}_{  \substack{ \II_3 \II_2 \\ \II_4 \II_5} } \(t_3,t_2;t_4\) =\gd_{\II_3,\II_4}\gd_{\II_2,\II_5}\gd\(t_3-t_4\)\gd\(t_4-t_2\)
 \\+\frac{\gd \Sigma_{\II_3 \II_2}\(t_3,t_2\)}{\gd G_{\II_6 \II_7}\(t_6,t_7\)}
G_{\II_6 \II_8}\(t_6,t_8\) \\ \times\wt{\Gamma}_{ \substack{\II_8 \II_9 \\ \II_4 \II_5}}\(t_8,t_9;t_4\) G_{\II_9 \II_7}\(t_9,t_7\).
\label{eq:map_Gamma_explicit}
\end{multline}
From \e{eq:map_Gamma_explicit} and \e{eq:4.8} follows the equation for the response function in the spinorial basis. Indeed
\begin{multline}
 \wt{L}_{  \substack{ \II_1 \II_{1'} \\ \II_2 \II_{2'}} } \(t_1,t_2\) =  G_{\II_1 \II_3}\(t_1,t_3\) \\ \times
 \wt{\Gamma}_{  \substack{ \II_3 \II_4 \\ \II_2 \II_{2'}} } \(t_3,t_4;t_2\)  G_{\II_4 \II_{1'}}\(t_4,t_1\).
 \label{eq:map_chi_explicit}
\end{multline}
It is crucial to observe that $\wt{L}$ is a two times and three space points function that can be contracted to define $\wt{\chi}$,
as explained in App.~\ref{app:X}.
Eqs.(\ref{eq:map_Dyson_explicit}--\ref{eq:map_chi_explicit}) represent the spinorial form of Hedin's equations.

\subsection{The $GW$ approximation}\label{sec:GW}
Starting from \e{eq:map_Sigma_explicit} and \e{eq:map_Gamma_explicit} the $GW$ approximation follows from chosing
\begin{equation}
\label{eq:gw.1}
\wt{\Gamma}_{  \substack{ \II_3 \II_2 \\ \II_4 \II_5} } \(t_3,t_2;t_4\) \approx \gd_{\II_3,\II_4}\gd_{\II_2,\II_5}\gd\(t_3-t_4\)\gd\(t_4-t_2\),
\end{equation}
from which
\eq{
\Sigma^{GW}_{\II\JJ}\(t_1,t_2\) =  -i G_{\LL\MM}\(t_1,t_2\) W^{RPA}_{  \substack{ \JJ \MM \\ \II \LL} }\(t_1,t_2\).
\label{eq:gw.2}
}
Starting from the $GW$ self--energy, different flavours of the scheme can be considered.
The $G_0W_0$ flavor assumes 
$G_{\II\JJ}(t_1,t_2)\sim G^{KS}_{\II\JJ}(t_1,t_2)$ and also $W^{RPA}$ functional of the KS states only.

From this point up to the end of the present section we will use the extended form of the spinor indexes, $\II\rar \(n\kk\)$ together
with the translational invariance.
Thus we assume to be in a perfectly periodic system, where the Coloumb interaction and the response function  are represented as a Fourier expansion
in terms of plane--waves, $\GG$, and transferred momenta, $\qq$.
\begin{subequations}
\label{eq:interaction_q_G}
\begin{align}
v\(\xx-\xx'\)=\sum_{\GG}\int \frac{d \qq}{\(2\pi\)^3} \frac{4 \pi}{|\qq+\GG|^2} e^{i\(\qq+\GG\)\cdot\(\xx-\xx'\)},
\label{eq:v_coulomb_q_G}
\end{align}
and
\begin{multline}
W^{RPA}\(\xx,\xx';\go\)= \sum_{\GG_1,\GG_2}\int \frac{d \qq}{\(2\pi\)^3} W^{RPA}_{\GG_1\GG_2}\(\qq;\go\) \\ \times
e^{i\(\qq+\GG_1\)\cdot\xx} e^{-i\(\qq+\GG_2\)\cdot\xx'}.
\label{eq:W_RPA_q_G}
\end{multline}
\end{subequations}
with $W^{RPA}\(\xx_1,\xx_2;\go\)$ the Fourier transform of $W^{RPA}\(1,2\)$.
The extended forms of \e{eq:interaction_q_G} can be found in several references, see for example Ref.~\onlinecite{Onida2002}.
By using \e{eq:W_RPA_q_G} we finally get that:
\ml{
\Sigma^{G_0W_0}_{nm\kk}\(\go\) = 
 -i \int \frac{d \go'}{2\pi}\int\,\frac{d^3\qq}{\(2\pi\)^3}\\
 \sum_{ij\GG_1 \GG_2} G^{KS}_{ij\kk-\qq}\(\go-\go'\)   
 W^{RPA}_{\GG_1\GG_2}\(\qq,\go'\)\\ 
\times \gr_{n i \kk}^{\qq}\(\GG_1\) \gr_{j m \kk}^{\qq,*}\(\GG_2\).
\label{eq:gw.5}
}
with 
\begin{align}
  \gr_{nm\kk}^{\qq}\(\GG\)= \sum_s  \int d\xx\, \phi_{n \kk s}\(\xx\)  \phi^*_{m \kk-\qq s}\(\xx\) e^{i\(\qq+\GG\)\cdot\xx}.
\label{eq:gw.6}
\end{align}
The Fourier transform of $G^{KS}_{nm\kk}(t_1,t_2)$ can be conveniently written as
\ml{
G^{KS}_{nm\kk}\(\go\)=\gd_{n,m}\\\[ \frac{\(1-f_{n\kk}\)}{\go-\gee^{KS}_{n\kk}+i O^+} + \frac{f_{n\kk}}{\go-\gee^{KS}_{n\kk}-i O^+}\].
\label{eq:dft.4}
}
We have now all ingredients to calculate the self--energy. Indeed, thanks to the definition \e{eq:gw.6}, all can be recast in the product of simple oscillators
that can be efficiently calculated via Fast Fourier Transormation techniques.\\

The use of the KS Hamiltonian as zero--th order term of the total Hamiltonian implies also that
$\Sigma_{nm\kk}\(\go\)$ needs to be replaced by $\Sigma_{nm\kk}\(\go\)-v^{xc}_{nm\kk}$ in the Dyson equation for $G$.
It follows then that Dyson equation reads
\ml{
G_{nm\kk}\(\w\)=G^{KS}_{nm\kk}\(\w\)+\\+G^{KS}_{ni\kk}\(\w\) \(\Sigma_{ij \kk}\(\w\)-v^{xc}_{ij\kk}\) G_{jm\kk}\(\w\).
\label{eq:gw.2a}
}
The last approximation in the $G_0W_0$ flavor is to assume that only the energies needs to be corrected and not the wave--functions.
This implies
\seq{
\eqg{
\Sigma_{nm\kk}\(\go\)\approx \gd_{nm}\Sigma_{nn\kk}\(\go\),\\
\label{eq:gw.1.3}
G_{nm\kk}\(\go\)\approx \gd_{nm}G_{nn\kk}\(\go\).
}
}
From \e{eq:gw.1.3} it follows the final form of Dyson equation, \e{eq:4.4} used in this work
\eq{
 \gee^{GW}_{n\kk}\approx \gee^{KS}_{n\kk} + (\Sigma_{nn\kk}\(\gee^{GW}_{n\kk}\)-v^{xc}_{nn\kk}).
 \label{eq:GW_energies}
}

\section{The Bethe--Salpeter equation}\label{sec:BSE}
The spinorial Bethe--Salpeter equation can be derived from the general spinorial Hedin's equations by some manipulations that we outline in the following.
Let's start by introducing the static limit of the $GW$ self--energy, the so called Screened Exchange\,($SEX$) approximation
\eq{
 \Sigma^{SEX}_{\II_1\II_2}\(t\) =  -i G_{\II_4\II_3}\(t\) W^{st}_{  \substack{ \II_1\II_4 \\ \II_3 \II_2} },
\label{eq:bse.1}
}
with $W^{st}=\gd(t_1-t_2)W^{RPA}\(t_1,t_2\)$. The approximation introduced by \e{eq:bse.1} is crucial in turning the BSE, an equation for a four point Green's function $\wt{L}$, 
in a simpler equation for a two time point function. Still the general solution is a four indexes function, $\wt{L}_{  \substack{ \II_1 \II_{1'} \\ \II_2
\II_{2'}} } \(\go\)$. 

We can now easily calculate the functional derivative $ \frac{\gd \Sigma_{\II_3 \II_2}}{\gd G_{\II_6 \II_7}} $ which defines the kernel of the BSE
for the vertex function:
\eq{
\frac{\gd \Sigma_{\II_3 \II_2}\(t_3,t_2\)}{\gd G_{\II_6 \II_7}\(t_6,t_7\)}\approx\gd\(t_3-t_6\)\gd\(t_2-t_7\) W^{st}_{\substack{ \II_3\II_6 \\ \II_7 \II_2} },
\label{eq:bse.2}
}
where we have assumed the derivative of $W^{st}$ to be negligible.
Thanks to the approximation \e{eq:bse.1} the equation of motion for the vertex acquires a simple form that can be solved in subspace of single frequency vertex
functions, $\Gamma\(\go\)$. From \e{eq:bse.2} and by using \e{eq:gw.1.3} we can work out the BSE for the spinorial vertex in the $SEX$ approximation:
\begin{multline}
\Gamma_{  \substack{ \II_3 \II_2 \\ \II_4 \II_5} } \(\go\) =\gd_{\II_3,\II_4}\gd_{\II_2,\II_5}
 +i\[ W^{st}_{  \substack{ \II_3 \II_6 \\ \II_7 \II_2} } - V_{  \substack{ \II_3 \II_6 \\ \II_7 \II_2} }\]
 \\ \times G_{\II_6\II_6}\(\go\)\wt{\Gamma}_{ \substack{\II_6 \II_7 \\ \II_4 \II_5}}\(\go\) G_{\II_7\II_7}\(\go\).
\label{eq:bse.3}
\end{multline}
In order to connect \e{eq:bse.3} to an equation of motion for the response function, the BSE, we now move from the $\II$ basis to the explicit $\(n\kk\)$
presentation. We start by introducing, for the general representation of $\wt{L}_{  \substack{ \II_1 \II_{1'} \\ \II_2 \II_{2'}} } \(t\)$ evaluated for a
given transferred momentum, $\qq$:
\seq{
\label{eq:bse.4}
\eqg{
 \II_1\equiv\(n\kk\),\\
 \II_{1'}\equiv\(n'\kk-\qq\),\\
 \II_{2}\equiv\(m\pp\),\\
 \II_{2'}\equiv\(m'\pp-\qq\).
}
}
We denote as $\wt{L}_{ \substack{nn'\kk\\mm'\pp}}\(\qq,t\)$ the response function whose scattering geometry is defined by \e{eq:bse.4}.
\e{eq:bse.3} now defines an equation for $\wt{L}_{ \substack{nn'\kk\\mm'\pp}}\(q,\go\)$ as, following the notation \e{eq:bse.4}, we can write:
\eq{
 \wt{L}_{ \substack{nn'\kk\\mm'\pp}}\(\qq,\go\)=G_{n\oo{n}\kk}\(\go-\oo{\go}\) G_{n'\oo{n}'\kk-\qq}\(\oo{\go}\) \Gamma_{  \substack{  \oo{n}\oo{n}'\kk\\mm'\pp} } \(\go\).
\label{eq:bse.5}
}
By putting together \e{eq:bse.5} and \e{eq:bse.3} we get the final equation for $\wt{L}$:
\ml{
L{  \substack{nn'\kk\\mm'\pp} } \(\qq,\go\) =
L^0_{ \substack{ nn'\kk\\mm'\pp }} \(\qq,\go\)+\\
 L^0_{ \substack{ nn'\kk\\\oo{n}\oo{n}'\oo{\kk} }} \(\qq,\go\)
  K_{\substack{\oo{n}\,\oo{n}'\oo{\kk}\\\oo{m}\,\oo{m}'\oo{\pp}}}(\qq)
     \wt{L}_{ \substack{\oo{m}\,\oo{m}'\oo{\pp}\\mm'\pp}}\(\qq,\go\)
\label{eq:bse}
}
with
\begin{eqnarray}
 L^0_{ \substack{ nn'\kk\\mm'\pp }} \(\qq,\go\) \equiv&& \gd(\kk-\pp) \gd_{nm}\gd_{n'm'}\times \nonumber \\
           && G_{nn\kk}\(\go-\oo{\go}\) G_{n'n'\kk-\qq}\(\oo{\go}\) \\
 -iK_{\substack{nn'\kk\\ mm'\pp}}(\qq)\equiv&&
   \[ W^{st}_{ \substack{nm\kk\\ n'm'\kk-\qq} }\(\kk-\pp\)- V_{\substack{nn'\kk\\ mm'\pp} }\(\qq\) \]
\label{eq:bse.7}
\end{eqnarray}
${\int\frac{\,d^3\oo{\kk}}{\(2\pi\)^3}}$, ${\int\frac{\,d^3\oo{\pp}}{\(2\pi\)^3}}$  and ${\int\frac{\,d\oo{\go}}{\(2\pi\)}}$ are implicit.
\e{eq:bse} is the  spinorial Bethe--Salpeter equation written for a generic transferred momentum $\qq$.
In the $\qq\rar\bs{0}$\,(optical) limit it reduces to the optical
BSE that we will use from now on to study optical properties in the next section. 

Like in the $G_0W_0$ case \e{eq:bse} looks the same of the scalar BSE. 
Indeed the only difference is the definition of the oscillators defined in \e{eq:gw.6}.
It also follows that, like in the spin independent case, the solution of \e{eq:bse} can be recast in an eigenvalue problem.
In order to show this it is enough to solve
\e{eq:bse} by noting that $L^0\(\go\)$ is a sum of simple single--pole functions.
Carefully separating the resonant and the anti--resonant term the eigenvalue problem
can be defined
\begin{eqnarray}
H^{exc} A^\gl = E_{\gl} M A^\gl  
\end{eqnarray}
in terms of an excitonic matrix $H^{exc}$ \cite{Onida2002} and a metric tensor
\begin{equation}
M=\begin{pmatrix}
1 &  0 \\
0 & -1
\end{pmatrix}
\end{equation}
The final form of ${L_{\substack{nn'\kk\\mm'\pp} }(\go)}$ can be expressed in terms of the eigenstate of the BSE matrix.
It has a particolar simple expression if the resonant only constributon is considered:
\begin{equation}
 L_{  \substack{nn'\kk\\mm'\pp} } \(\go\)=
\(
\sum_{\gl} \frac{ A^{\gl,*}_{nn'\kk} A^{\gl}_{mm'\pp} } { \go -E_{\gl}+i0^{+}}
\)
\label{eq:bse.8}
\end{equation}
Starting from the eigen--vectors of the excitonic hamiltonian, we can define the excitonic state wave--function as
a linear combination of electron--hole pairs:
\begin{equation}
\label{eq:exc_spin_1}
\olrar{|\gl\ra}=\sum_{nm\kk} A^\gl_{nm\kk} \orar{ | n \kk \ra} \otimes \orar{| m \kk \ra} .
\end{equation}
The electron--hole pairs contributing to the excitonic wave--function are vectors in the spin space.
In practice this means that the exciton  is, in the non collinear case, a linear combination of
the four possible spin orientations of the electron and hole, i.e. a tensor. 

\subsection{Blocking of the BSE matrix and BSE spin tructure}\label{sec:BSE_blocking}
\subsubsection{The magnetic case}
If $\h{v}^{\(SOC\)}$ is weak, the non--collinearity of the KS eigenstates can be neglected
as first step and then the SOC correction treated perturbatively.
In the collinear limit case $s_z$, i.e. the spin projection of electrons (and holes), becomes a good quantum number
and accordingly also $S_z$, i.e. the spin projection of the exciton.
The extra quantum number can be added in \e{eq:spinor_def} which becomes
\begin{align}
\la \xx\orar{ | n \kk s \ra}\equiv \orar{\psi}_{\II s}\(\xx\)= \phi_{n\kk s}\(\xx\) \orar{|s\ra,}
\label{eq:spinor_def_coll}
\end{align}
where $\overrightarrow{|s\ra}$ is either ${\overrightarrow{|\up\ra}=(1,0)^t}$ (for $s=+1/2$) or ${\orar{|\dn\ra}=(0,1)^t}$ (for $s=-1/2$);
the superscript ``$t$'' indicates the transposition operation.
If a collinear calculation is performed, in practice one can compute just $\phi_{n\kk s}\(\xx\)$ and reconstruct the spinorial wave--functions via Eq.~\eqref{eq:spinor_def_coll}. This would be, in principle, also the result of a spinorial calculation with a collinear hamiltonian.
However in this second case, whenever $\epsilon_{n\kk \up}=\epsilon_{n\kk \dn}$,
the resulting wave--functions will be a random (but ortogonal) linear combination of
$\overrightarrow{\psi}_{\II \up}\(\xx\)$ and $\overrightarrow{\psi}_{\II \dn}\(\xx\)$ wave--functions.

In the collinear limit $H^{exc}$ can be blocked in two matricies
with half the size of the spinorial BSE:
the spin conserving  ($\Delta S_z=0$) transitions or ``excitons'' ($\gl_e$) and the spin flip  ($\Delta S_z=\pm 1$) transitions or ``magnons'' ($\gl_m$), with
\seq{
\label{eq:exc_spin_12}
\eqg{
|\gl_e\ra=\sum_{c v\kk s } A^{\gl_e}_{cv\kk s} |c\kk, s\ra \otimes |v\kk, s\ra,\\
|\gl_m\ra=\sum_{c v\kk s } A^{\gl_m}_{cv\kk s} |c\kk, s\ra \otimes |v\kk, - s\ra.
}
}
Excitons and magnons, in \e{eq:exc_spin_12}, distinguish the two possible spin combinations of the electron--hole pair.
In the collinear case, only excitons, where the $c$ and $v$ states have the same spin, can be excited by the laser pulse.
Magnons cannot be generated, since optical transitions between state with opposite spin are forbidden.
Notice that the magnons block is also composed by two independent subblocks which are in general different,
the $\Delta S_z=+1$ and the $\Delta S_z=-1$. These are two independent set of excitations when $S_z$ is a good quantum number. For an analysis of BSE applied to the magnon channel see for example Ref.~\cite{Muller2016}.
Instead the total spin is not yet a good quantum number and the origin of this can be traced back to the fact that
$|\phi_{n\kk\up}\(\xx\)|^2\neq |\phi_{n\kk\dn}\(\xx\)|^2$.
This is well known in the literature of quantum chemistry, where the term spin contamination is used and restricted calculations,
which indeed impose $\phi_{n\kk\up}\(\xx\)= \phi_{n\kk\dn}\(\xx\)$ and ${\epsilon_{n\kk \up}=\epsilon_{n\kk \dn}}$, are
sometimes performed. In extended systems the breaking of spin symmetry is instead regarded as less important; the exchange
splitting ${\Delta_{n\kk}=\epsilon_{n\kk \up}-\epsilon_{n\kk \dn}}$ is seen as a physical quantity.

\subsubsection{The non--magnetic case}
Instead if the system is non--magnetic ${\epsilon_{n\kk \up}=\epsilon_{n\kk \dn}}$
for any $n\kk$, the ground state has total spin $S=0$, and
\eq{
\phi_{n\kk s}\(\xx\)=e^{i\ga_{n\kk s}}\phi_{n\kk}\(\xx\),
\label{eq:exc_spin_16}
}
with $\ga_{n\kk s}$ an arbitrary phase factor.
In practical calculations only $\phi_{n\kk}\(\xx\)$ are computed and $\ga_{n\kk s}=0$ is assumed.
However, if a collinear spin dependent calculation is explicitly performed on a non magnetic systems,
random phases $\ga_{n\kk\uparrow}-\ga_{n\kk\downarrow}$ will be present in between the two spin channels.
Using \e{eq:exc_spin_16} the BSE Hamiltonian the exciton channel ($\Delta S_z=0$) can be further
blocked into singlets (S) with $(S,S_z)=(0,0)$ and triplets (T)
with $(S,S_z)=(1,0)$. The ``magnons'' channel remains composed of two bocks which now represent the triplets
$(S,S_z)=(1,+1)$ and  $(S,S_z)=(1,-1)$. All triplet blocks are degenerate and identical up to the phases $\ga_{n\kk}$.
Starting from the solutions of the unpolarized BSE $\tilde{A}^{\gl_S}_{cv\kk}$ ($\tilde{A}^{\gl_T}_{cv\kk}$)
in the singlet (triplet) block, the full eigenvectors can be reconstructed as
\seq{
\label{eq:exc_spin_17}
\eqg{
  A^{\gl_S}_{cv\kk s}= (-1)^{2s+1} \frac{e^{i\(\ga_{c\kk s}-\ga_{v\kk s}\)}}{\sqrt{2}} \tilde{A}^{\gl_S}_{cv\kk}, \label{eq:exc_spin_17a}\\
  A^{\gl_T}_{cv\kk s}= \ \ \ \ \ \ \ \ \ \ \ \frac{e^{i\(\ga_{c\kk s}-\ga_{v\kk s}\)}}{\sqrt{2}} \tilde{A}^{\gl_T}_{cv\kk} . \label{eq:exc_spin_17b}
}
}
The factor $\sqrt{2}$ in \e{eq:exc_spin_17} ensures that the eigenvectors are normalized to 1.

\subsubsection{BSE matrix blocking}
\label{sect_BSE_blocking}
To summarize let us explicitly write the general spin structure of the
matrix relabelling the states $n\kk$ (with $n=1...N$) as $\nt_s\kk$ (with $\nt=1...N/2$, $s=\up,\dn$). This is just an exact re-labelling, which becomes meaningful, i.e. ${\nt_\up\kk}$ (${\nt_\dn\kk}$) refers to a ``spin-up'' (``spin-down'') state in case ${\h{v}^{\(SOC\)}}$ is small and the collinear case notation can be recovered, i.e. ${\{\nt_\up\kk\}=\{m\kk\up\}}$ if ${\h{v}^{\(SOC\)}=0}$.
For each set of indexes ${\{\nt\nt'\kk,\mt\mt'\pp\}}$
\begin{equation}
H^{res}=H^0+H^{exch}+H^{eh-int}
\label{eq:Hexc_zero_and_int}
\end{equation}
is written in terms of $4\times 4$ matrices in the ``spin''
indexes~\footnote{An alternative and equivalent way would be to directly write the BSE in the basis set of the calculation without SOC. In such case the matrix elements of the SOC potential and of the difference $v^{Hxc}[\rho^{SOC}]-v^{Hxc}[\rho^{0}]$ would explicitly appear in $H^0$}:
\seq{
\label{eq:Hexc_zero_and_int_spin}
\begin{eqnarray}
H^0_{s_1s_2s_3s_4}&=&(\epsilon_{s_1}-\epsilon_{s_2})\gd_{s_1,s_3}\gd_{s_2,s_4},  \\
H^{exch}_{s_1s_2s_3s_4}&=&V_{s_1s_2s_3s_4}, \\
H^{eh-int}_{s_1s_2s_3s_4}&=&W^{st}_{s_1s_3s_2s_4}.
\end{eqnarray}
}

We have three cases:
\begin{itemize}
\item non collinear case.\\
Neither $S_z$ nor $S$ are good quantum numbers, thus excitons and magnons are mixed.
$H^{exc}$ is ${N\times N}$ matrix.
All the matrix elements of Eq.\eqref{eq:Hexc_zero_and_int_spin} can be different from zero. 

\item Collinear magnetic case.\\
$S_z$ is a good quantum number, while $S$ is not.
The BSE can be blocked in two matrices, $H^{e}$ and $H^{m}$, of size ${N/2\times N/2}$ each, which describe separately excitons and magnons.
This results from the fact that, for the collinear case, ${V_{s_1s_2s_3s_4}\propto \gd_{s_1,s_2}\gd_{s_3,s_4}}$ and
${W^{st}_{s_1s_3s_2s_4}\propto \gd_{s_1,s_3}\gd_{s_2,s_4}}$, and thus all matrix elements
which couple the $\Delta S_z=0$ channel and to the the $\Delta S_z=\pm 1$ channels are zero.

\item Collinear non-magnetic case.\\
Both $S_z$ and $S$ are good quantum numbers.
The BSE can be further blocked with four blocks in total three of which carry the same information.
The exciton channel $H^e$ generates the singlets block ($\Delta S=0$) and the triplets block ($\Delta S=1$).
The two blocks resulting from the magnon channel $H^m$ are equivalent to the triplet block.
Indeed, since the ground state is non-magnetic, the distinction between excitons and magnons becomes meaningless.  
In total there are two matrices, $H^{S/T}$, of size ${N/4\times N/4}$ each. This results from the fact that $V_{s_1s_1s_3s_3}= V$
for any $\{s_1,s_3\}$ and ${W^{st}_{s_1s_1s_2s_2}=W^{st}}$ for any $\{s_1,s_2\}$.
The blocking in this second step is obtained via the vectors $1/\sqrt{2}(1,1)^t$ and $1/\sqrt{2}(1,-1)^t$.
As a consequence $H^{exch,S}=2V$ while $H^{exch,T}=0$ and all triplet states are lower in energy.
\end{itemize}

\section{The Excitonic spin polarization}\label{sec:BSE_spin}
In the previous section we have discussed the spin structure of $H^{exc}$ in different cases and connected the different sized of the matrix to the spin of the exciton in the singlets and triplets channels. We start the present section
introducing the matrix which relates the spin of the electron and the hole to the total spin $S$ and its projection $S_z$ of the electron--hol pair. We define the triplet and singlet spin states in the usual way, $|S,S_z\ra$~\cite{Sakurai1994},
\eq{
|S,S_z\ra = R^{SS_z}_{s_e s_h} |s_e\ra \otimes |s_h\ra.
\label{eq:exc_spin_4}
}
$R^{SS_z}_{s_es_h}$ are matrices in the basis of the products of the electron and hole spins. These can be written as 
2$\times$2 matrices as follows
\begin{subequations}
\label{eq:exc_spin_5}
\eqg{
 \uu{\uu{R}}^{1,-1}=
 \(
  \begin{array}{cc}
    0 & 0 \\ 1 & 0
  \end{array}  
  \),\\
 \uu{\uu{R}}^{1,0}=\frac{1}{\sqrt{2}}
 \(
  \begin{array}{cc}
    -1 & 0 \\ 0 & 1
  \end{array}  
  \),\\
 \uu{\uu{R}}^{1,1}=
 \(
  \begin{array}{cc}
    0 & 1 \\ 0 & 0
  \end{array}  
  \),\\
 \uu{\uu{R}}^{0,0}=\frac{1}{\sqrt{2}}
 \(
  \begin{array}{cc}
    1 & 0 \\ 0 & 1
  \end{array}  
  \).
}
\end{subequations}

We start by expanding \e{eq:exc_spin_1} in the basis of space/spin components and write explicitly the $s_e s_h$ component of the excitonic state:
\eq{
\label{eq:exc_spin_2}
|\gl\ra=  \sum_{s_e s_h}\int d^3\rr_e d^3\rr_h 
|s_e s_h\ra |\rr_e \rr_h\ra 
\la \rr_e \rr_h| \la s_e s_h| \gl\ra.
}
\e{eq:exc_spin_2} defines the excitonic wavefunction projected on the spin state $|s_e s_h\ra |\rr_e \rr_h\ra$:
\eq{
 \Psi_{s_e s_h}^{\gl}\(\rr_e \rr_h\)\equiv \la \rr_e \rr_h| \la s_e s_h| \gl\ra.
\label{eq:exc_spin_3}
}
Thanks to \e{eq:exc_spin_4} we can rewrite \e{eq:exc_spin_3} in terms of components on the triplet and singlet spin states
\eq{
\label{eq:exc_spin_6}
|\gl\ra= \sum_{S S_z} \int d^3\rr_e d^3\rr_h 
|S,S_z\ra |\rr_e \rr_h\ra 
\la \rr_e \rr_h| \la S,S_z| \gl\ra,
}
with
\ml{
\Psi_{S,S_z}^{\gl}\(\rr_e \rr_h\)\equiv \la \rr_e \rr_h| \la S,S_z| \gl\ra=\\\sum_{s_e s_h} R^{S,S_z}_{s_e s_h}  \Psi_{s_e s_h}^{\gl}\(\rr_e \rr_h\)
\label{eq:exc_spin_7}
}
Thanks to \e{eq:exc_spin_7} we can introduce several observables that can efficiently describe the non--collinerarity of the excitonic state.

We start from the normalization condition:
\ml{
\gd_{\gl \gl'}= \la \gl | \gl' \ra =  \sum_{S,S_z} \int d^3\rr_e d^3\rr_h |\Psi_{S,S_z}^{\gl}\(\rr_e \rr_h\)|^2=\\
\sum_{S,S_z} N^{\gl}_{S,S_z},
\label{eq:exc_spin_8}
}
with
\ml{
 N^{\gl}_{S,S_z}\equiv |\la S,S_z | \gl \ra|^2=  \int d^3\rr_e d^3\rr_h\\
 \Bigl\lvert \sum_{c v \kk s_e s_h }  A^{\gl}_{cv\kk}  R^{S,S_z}_{s_e s_h} \phi_{c \kk s_e}\(\rr_e\) \phi^*_{v \kk s_h}\(\rr_h\)\Bigr\rvert^2.
\label{eq:exc_spin_9}
}
It is worth noting that when expanding the square modulus of \e{eq:exc_spin_9}, only a single summation over 
$\kk$ survives due to momentum conservation, whereas on all the other indexes double summations remain. 
From \e{eq:exc_spin_9} it follows that we can define an excitonic total average spin and momentum as
\seq{
\label{eq:exc_spin_10}
\eqg{
\label{eq:exc_spin_10.1}
 S^2_{\gl} = \la \gl | \h{S}^2 | \gl \ra=\sum_{S_z} S\(S+1\) N^{\gl}_{S,S_z},\\
\label{eq:exc_spin_10.2}
 S_{z,\gl} = \la \gl | \h{S}_z | \gl \ra= \sum_S S_z N^{\gl}_{S,S_z}.
}
}

\subsection{The collinear magnetic case}\label{sec:EXC_SPIN_collinear_magn}
The meaning of \e{eq:exc_spin_9} can be better understood by taking the collinear case. 
From \e{eq:exc_spin_9} it follows that we can define the spin polarization of excitons and magnons as:
\ml{
\label{eq:exc_spin_13}
	N^{\gl_{e,m}}_{S,S_z}=
 \sum_{\substack{cv\\c'v'}\kk} \sum_{ss'} A^{\gl_{e,m},*}_{c'v'\kk s'}  A^{\gl_{e,m}}_{cv\kk s } \\
  R^{S,S_z,*}_{s \pm s}  R^{S,S_z}_{s' \pm s'}  \la c' \kk, s' | c \kk, s \ra  \la v \kk, \pm s | v' \kk, \pm s' \ra,
}
where the $+$ and $-$ signs stand for the excitons and magnons channel respectively.
Note that in \e{eq:exc_spin_13} the two inner products are between the spatial part of the wave--function, which does not impose the spin to be conserved.

In order to manipulate \e{eq:exc_spin_13} we observe, from \e{eq:exc_spin_5}, that the specific form of the $R$ matrices impose that in the $N^{\gl_e}_{S,S_z}$
case only the $\(S,S_z\)=\(1,0\)$ and $\(0,0\)$ are
non zero. On the contrary in the $N^{\gl_m}_{S,S_z}$ case the non--zero components will be $\(S,S_z\)=\(1,\pm 1\)$.
This marks the distinction between excitons and magnons.

After some simple manipulation, we get
\seq{
\label{eq:exc_spin_14}
\begin{eqnarray}
N^{\gl_e}_{S0}&=& \frac{1}{2}\wl 1\pm \sum_{\substack{cv\\c'v'}\kk} \[O_{\substack{cv\\c'v'}\kk \up\dn}+O_{\substack{cv\\c'v'}\kk \dn\up}\] \\
 O_{\substack{cv\\c'v'}\kk ss'} &=& A^{\gl_e,*}_{c'v'\kk s}  A^{\gl_e}_{cv\kk s'} \times \nonumber\\
                               && \ \ \ \ \la c' \kk, s | c \kk, s' \ra  \la v \kk, s' | v' \kk, s\ra ,
\end{eqnarray}
}
where the $+$\,($-$) refers to the $S=0$\,($S=1$).  Similarly we get
\eqg{
N^{\gl_m}_{1 1}=  \sum_{cv\kk} |A^{\gl_m}_{cv\kk \up}|^2,\\
N^{\gl_m}_{1-1}=  \sum_{cv\kk} |A^{\gl_m}_{cv\kk \dn}|^2.
\label{eq:exc_spin_15}
}
From \e{eq:exc_spin_14} and \e{eq:exc_spin_15} it follows that, in general, also in the collinear case the value of $S_\gl$ is not fixed. 
This is again a manifestation of the fact that $S$ is not a good quantum number in general. The collinear systems can be characterized by a non vanishing magnetization that makes the $\up$ and $\dn$ components of the electronic wave--functions to differ by more than a simple phase factor. The consquence is a state--dependent value of the excitonic spin.

\subsection{The collinear non--magnetic case}\label{sec:EXC_SPIN_collinear_non_magn}
Let's conclude this section by considering the case of a collinear and non--magnetic system. 
When \e{eq:exc_spin_16} and \e{eq:exc_spin_17} are plugged in \e{eq:exc_spin_13} it turns out that
\seq{
\label{eq:exc_spin_18}
\eqg{
 N^{\gl_e(S)}_{S S_z}=\gd_{S,0} \gd_{S_z,0},\\
 N^{\gl_e(T_0)}_{S S_z}=\gd_{S,1} \gd_{S_z,0},\\
 N^{\gl_m(T_{1})}_{S S_z}=\gd_{S,1} \gd_{S_z,-1}, \\
 N^{\gl_m(T_{-1})}_{S S_z}=\gd_{S,1} \gd_{S_z,1}.
}
}
as it should from the blocking of the BSE matrix.

\section{A perturbative approach to the effect of Spin--Orbit coupling}\label{sec:pert_SOC}
We now proceed considering the SOC as a perturbation.
To first order the perturbation just gives a correction to the DFT eigenvalues and does not touch the wave--functions.
Numerically the perturbation need to be applyed by mapping the solution of the DFT calculation without SOC into the solution of the DFT 
calculation with SOC. In the first part of this section we discuss such mapping. 

Once a mapping is defined, the SOC perturbation could be directly applied to the KS energies before solving both the GW scheme and the BSE
or after. In the second part of this section we discuss such distinction.

\subsubsection{Mapping procedure}
Now, in order to define a mapping procedure in practice we distinguish in between the KS hamiltonian without SOC,
$\h{h}^{KS,0}$, and the standard KS hamiltonian $\h{h}^{KS}$, with  $\h{h}^{KS}-\h{h}^{KS,0}=\h{v}^{\(SOC\)}$.
Moreover, let us focus here on the situation where the ground state is non--magnetic, since it is the case for which calculations
are actually performed in the present manuscript and also because it is the more complex case. The generalization to magnetic
systems is straightforward.
$\h{h}^{KS,0}$ is then collinear and spin independent. It has
eigenvectors $\phi^0_{n\kk}\(\xx\)$ and energies $\gee^{KS,0}_{n\kk}$. 

The key passage now is how to connect the eigenvectors of SOC free case to the full spinorial case.
To this end we define a map based on the overlap between $\phi^0_{n\kk}\(\xx\)$ and $\orar{\phi}_{n\kk}\(\xx\)$.
Since both represent a complete basis--set for each k--point,
we just need to expand $\orar{\phi}_{n\kk}\(\xx\)$ in terms of the spinor defined
extending $\phi^0_{m\kk}\(\xx\)$ first including the spin index (i.e. using Eq.~\eqref{eq:exc_spin_16}),
and then to constructing the spinors (i.e. using Eq.~\eqref{eq:spinor_def_coll}).
Notice that in doing so the gauge with $e^{is\alpha_{n\kk}}=1$ is assumed.
Moreover, in the degenerate spaces, the ``up'' and ``down''  spinors
are always chosen among all possible random combinations.
Let as call $\orar{\phi}^0_{m\kk s}\(\xx\)=\langle  \xx | \orar{m\kk s} \rangle$ the result of such expansion.
Then it follows
\eq{
|\orar{n\kk}\rangle = \sum_{ms} | \orar{m\kk s} \rangle \langle \olar{m\kk s} | \orar{n\kk} \rangle  
                           = \sum_{ms} \Delta^{\kk}_{n,ms} | \orar{m\kk s}\rangle,
 \label{eq:pert_soc_map}
}
We now want compute the first order perturbation theory correction to the energy
with respect to the perturbation $\hat{V}=\h{h}^{KS}-\h{h}^{KS,0}$. Using Eq.~\eqref{eq:pert_soc_map} it can be expressed as
\begin{eqnarray}
	\epsilon_{m\kk s} &=& \epsilon^0_{m\kk} + \langle \olar{m\kk s} | (\h{h}^{KS}-\h{h}^{KS,0}) | \orar{m\kk s} \rangle \nonumber \\
&=& \epsilon^0_{m\kk} + \(\langle \olar{m\kk s} | \h{h}^{KS} | \orar{m\kk s} \rangle - \epsilon^0_{m\kk} \) \nonumber \\
&=& \epsilon^0_{m\kk} + \sum_{n} |\Delta^{\kk}_{n,ms}|^2 \epsilon_{n\kk}-\epsilon^0_{m\kk} \label{eq:pert_soc_def}
\end{eqnarray}
where in the third line we inserted two completness relations over $|\orar{n\kk}\rangle$, used the fact that $|\orar{n\kk}\rangle$ are eigenstates of $\h{h}^{KS}$, and used the definition of $\Delta^{\kk}_{n,ms}$.
Eq.~\eqref{eq:pert_soc_def} is exact.
We now define a generic mapping function, $f_\kk$, which maps every eigenstate of $\h{h}^{KS}$ to the eigenstates of $\h{h}^{KS,0}$.
In case neither $|\orar{m\kk s}\rangle$ nor $|\orar{n\kk}\rangle$ are degenerate,
the mapping function can be defined as
\begin{equation}
	f_\kk(ms)=n \quad\text{if}\quad  |\Delta^{\kk}_{n,ms}|^2= D^\kk_n,
\end{equation}
with ${D^\kk_n=\max_{ms}|\Delta^{\kk}_{n,ms}|^2}$.
The extension to the degenerate case is discussed is app.~\ref{App:mapping_degenerate}.
Using the mapping procedure we can approximate Eq.~\eqref{eq:pert_soc_def} as
\begin{eqnarray}
\epsilon_{m\kk s} 
&\approx& \epsilon^0_{m\kk} + (\epsilon_{f_\kk(ms)\kk}-\epsilon^0_{m\kk}) \nonumber \\
&\approx& \epsilon^0_{m\kk} + \Delta \epsilon^{SOC}_{m\kk s}
\end{eqnarray}
Note also that for each  eigenvalue $\epsilon^0_{m\kk}$ a pair of spinorial state is selected, one for $s=\up$ and one for $s=\dn$, since to each $m\kk$ are associated two spinors via Eq.~\eqref{eq:exc_spin_16}) and Eq.~\eqref{eq:spinor_def_coll}).
This property will be crucial to define spin-conserving ($\Delta S_z\approx 0$) and spin-flip ($\Delta S_z\approx \pm 1$) excitations.

We also define the quality of the mapping as
\begin{equation}
Q = \min_{n\kk} \( D^\kk_n \),
\end{equation}
and we monitor its value. $0<Q\leq 1$, and $Q=1$ means the mapping is exact.

\subsubsection{Perturbative SOC within GW and BSE}
We now want to use the mapping to apply the corrections in the GW-BSE scheme.
Formally one should first apply the SOC corrections and after solve
the GW-BSE scheme. Applying them before would significantly increase the computational load, thus reducing
the advanges of using a perturbative approach, compared to the full approach.
We thus want to apply the corrections after. In the next section we will check the quality of the scheme against full SOC 
calculations.

Both in GW and in BSE the KS energies enter in the $RPA$ screening.
In $W^{RPA}$ one can reasonably expect no significant changes due to SOC.
Then, as far as GW is concerned, the KS energies enter in two more points: the definition of the self--energy
and the Dyson equation for $G$.
The Dyson equation for $G$ is recast in terms of Eq.~\eqref{eq:GW_energies} for the QP energies
and gives
\begin{equation}
 \gee^{GW}_{m\kk s} \approx \gee^{KS}_{m\kk} + (\Sigma_{mm\kk}\(\gee^{GW}_{m\kk}\)-v^{xc}_{mm\kk}) + \Delta \epsilon^{SOC}_{m\kk s}.
 \label{eq:GW_energies_SOC}
\end{equation}
Since the perturvative approach assumes only the energies change, all matrix elements remain in the collinear basis set. Only the self--energy depends explicitly on the energies. Thus small differences in between applying SOC corrections before or after can be due to the self-energy. Further differences with the full SOC approach are expected to be higher order.

In the BSE scheme instead there is no strightforward way to apply the corrections after and a new procedure needs to be defined.
To this end we follow the ``perturbative--BSE'' (pBSE) approach of Ref.~\onlinecite{Qiu2013}, which however we critically discuss and refine.
In the pBSE the SOC potential extendend to the eh--basis set is defined
\begin{equation}
V^{\(SOC\)}_{ss'}\(\xx_c,\xx_v\) = (v^{\(SOC\)}_{ss',c}\(\xx_c\) \otimes \mathds{1}_v) - ( \mathds{1}_c \otimes v^{\(SOC\)}_{ss',v}\(\xx_v\)).
\label{eq:V_SOC_eh}
\end{equation}
This potential is then used to define the SOC corrections $\Delta\omega_{\lambda s}$ to the BSE eigenvalues
without SOC $\omega^0_\lambda$ via its expectation value on the excitonic state.

For the spin conserving, optically active, channel we use the mapping procedure proposed in the literature~\onlinecite{Qiu2013}. Introducing
\begin{eqnarray}
\Psi^{\lambda S,0}_{ss'}(\xx_c,\xx_v)&=&\delta_{s,s'}\sum_{cv\kk s}A^{\lambda S}_{cv\kk s}\phi^{0,*}_{c\kk s}(\xx_c)\phi^0_{v\kk s}(\xx_v) \nonumber
\end{eqnarray}
where $A^{\lambda S}_{cv\kk s}$ is obtained from eqs.~\eqref{eq:exc_spin_17}
the SOC corrections are defined as:
\begin{eqnarray}
\Delta\omega_{\lambda}^{S,0} &=& \langle \Psi^{0,\lambda} | \h{V}^{\(SOC\)} | \Psi^{0,\lambda} \rangle \nonumber \\
&=& \sum_{cv\kk s}\sum_{c'v'\kk' s'}  A^{\lambda S,*}_{cv\kk s} A^{\lambda S}_{c'v'\kk' s} \times \nonumber \\
&& \ \ \ \ \ \Big[ \langle \psi^0_{c\kk s} | \h{v}^{\(SOC\)} | \psi^0_{c'\kk' s'} \rangle + \nonumber \\
&& \ \ \ \ \ \ \ \ \ \ \ \  \langle \psi^0_{v\kk s} | \h{v}^{\(SOC\)} | \psi^0_{v'\kk' s'} \rangle  \Big] \nonumber \\
&\approx& \sum_{cv\kk s} |A^{\lambda S}_{cv\kk s}|^2 (\Delta \epsilon^{SOC}_{c\kk s}-\Delta \epsilon^{SOC}_{v\kk s}) \label{eq:pert_soc_singlets_Qui2013}
\end{eqnarray}
We used the mapping approximation and we neglect the terms with $c\neq c'$, $v\neq v'$ and $s\neq s'$. $\gd(\kk-\kk')$ is instead imposed by the Bloch hamiltonians. 
This is a simplification and one should carefully check how the SOC splitting compares with the exchange splitting (see App.~\ref{App:SOC_and_exchange} for more details)

For the spin flipping (or magnons) channel instead the exchange interaction is always $\approx 0$ and the mapping procedure is unique, with
\begin{eqnarray}
\Psi^{\lambda T,\pm 1}_{ss'}(\xx_c,\xx_v)&=&\delta_{s,-s'}\sum_{cv\kk}A^{\lambda T}_{cv\kk s}\phi^{0,*}_{c\kk \pm s}(\xx_c)\phi^0_{v\kk \mp s}(\xx_v), \nonumber
\end{eqnarray}
and
\begin{eqnarray}
\Delta\omega_{\lambda}^{T,\pm 1} \approx \sum_{cv\kk } |A^{\lambda T}_{cv\kk s}|^2 (\Delta \epsilon^{SOC}_{c\kk \pm s}-\Delta \epsilon^{SOC}_{v\kk \mp s}) 
\label{eq:pert_soc_triplets}
\end{eqnarray}
Eqs.~\eqref{eq:pert_soc_singlets_Qui2013} and \eqref{eq:pert_soc_triplets} are used in the present manuscript to compute the optical properties of TMDs.

\section{Results for paradigmatic materials}\label{sec:RES}
Bulk TMDs are indirect gap semiconductors, but when
going to a single layer, their gap becomes direct making them
suitable for applications in the fields of electronics, optoelectronics and sensing.
The K$^+$ and K$^-$ points of
the hexagonal Brillouin zone are the location of TMDs' band extrema.
At these inequivalent points, linked by time-inversion symmetry, the
spin-split valence band maximum (VBM) and conduction band minimum
(CBM) are almost completely spin-polarized in an opposite way in the
two valleys, allowing for a selective valley excitation by $\sigma^{+}$
and $\sigma^{-}$ polarized light. 
All the results that will be discussed from this point were
obtained using the computational methods and parameters presented in App. \ref{App:compdetails}.

\begin{table}[!htbp]
\begin{tabular}{|l|c|c||c|c||c|c|}
\hline
\multirow{3}{*}{} 
  & \multicolumn{6}{c|}{SOC splitting} \tabularnewline
\cline{2-7} 
  & \multicolumn{4}{c||}{Present work} &   \multicolumn{2}{c|}{Literature}  \tabularnewline
\cline{2-7} 
  & \multicolumn{2}{c||}{VBM} & \multicolumn{2}{c||}{CBM } & \multicolumn{2}{c|}{CBM} \tabularnewline
\cline{2-7} 
 & DFT  & GW  & DFT  & GW  & DFT  & GW  \tabularnewline
\hline 
WSe$_{2}$  & 458  & 516  & 46 & 25 & 37 \cite{Echeverry2016} 40 \cite{zhang_heinz_natnano2017} & 7\cite{Druppel_PRB2018}, 10 \cite{Thorsten_PhysRevB.96.201113},
6 \cite{Echeverry2016}\tabularnewline
WS$_{2}$  & 408  & 394  & 33 & 12 & 33 \cite{Echeverry2016}  & 10 \cite{Druppel_PRB2018}, 12 \cite{Thorsten_PhysRevB.96.201113},
5 \cite{Echeverry2016}\tabularnewline
MoSe$_{2}$  & 186  & 191  & -23 & -29 & -21 \cite{Echeverry2016}  & -42 \cite{Druppel_PRB2018}, -41 \cite{Thorsten_PhysRevB.96.201113},
-14 \cite{Echeverry2016}\tabularnewline
MoS$_{2}$  & 145  & 151  & -3 & -9 & -3 \cite{Echeverry2016}  & -15 \cite{Druppel_PRB2018}, -15 \cite{Thorsten_PhysRevB.96.201113},
-31 \cite{Echeverry2016}\tabularnewline
\hline 
\end{tabular}\caption{SOC splitting of the highest valence and lowest conduction band (VBM, CBM) at K. 
The conduction band splitting is defined as: $CBMS=\epsilon_{CBM}^{\sigma}-\epsilon_{CBM}^{-\sigma}$. 
Where $\sigma$ is the spin of the top of the valence band at K.
All energies are in meV. \label{tab:band_splittings}}
\end{table}

\subsection{Effect of SOC in the band structure}
As shown in Tab. \ref{tab:band_splittings}
the valence band spin splitting at K is of several hundreds of meV,
whereas the conduction band is splitted only few tens of meV. 
SOC splitting of the highest valence and lowest conduction band (VBM, CBM) at K.
Because the lowest SOC split conduction band can either have the same or the opposite 
spin character of the top valence band, it is convenient to define 
the conduction band splitting as: $CBMS=\epsilon_{CBM}^{\sigma}-\epsilon_{CBM}^{-\sigma}$,
where $\sigma$ is the spin of the top of the valence band at K. In this way the two different
situations are marked by the sign of the splitting.

Similarly to previous calculations, and shown in Figs. \ref{fig:mo_based_bands_at_k} and 
\ref{fig:w_based_bands_at_k}, we observe that for Mo-based systems the 
bottom of conduction at K has the same character of
the top of the valence whereas the opposite is
true for W-based ones. We find a reduction (increase) of the  
absolute value of the CBM splitting for the WX$_2$ (MoX$_2$) systems, as
the conduction state with the same spin character of the VBM 
is less corrected than the one with the opposite character 
(if $\sigma$ is the spin of the top of the valence band at K, $\Delta \epsilon^{GW}_{\sigma,c}< \Delta \epsilon^{GW}_{-\sigma,c}$).
At the GW level, the sign of the CB splitting is consistent with the literature,
the observed dispersion in the computed numerical values is possibly due to the use 
of the GdW approach in one case \cite{RohlfingPRB2010} and to different
calculation parameters or numerical implementations in the remaining cases 
\cite{Echeverry2016,zhang_heinz_natnano2017,Druppel_PRB2018,Thorsten_PhysRevB.96.201113}. 

As a consequence of the positive/negative CBM splitting, based only on the analysis
of the GW electronic states, it could be argued that for W-based systems
the optically active and spin allowed transition should have higher
energy with respect to the dark spin-forbidden transition, while the
opposite should be true for Mo-based ones. 
However it is not possible to predict whether the lowest optical excitation 
is spin-forbidden (dark) or spin-allowed (bright) only on the basis of the
electronic bandstructure. The dark or bright character of the
lowest optical excitation is indeed the result of a delicate interplay
between the different contributions to the excitation energy and the
full treatment of SOC at a non perturbative level, may change qualitatively the picture. 

\subsection{Optical absorption spectra}
\begin{figure*}[!hbt] 
\includegraphics[width=0.8\textwidth]{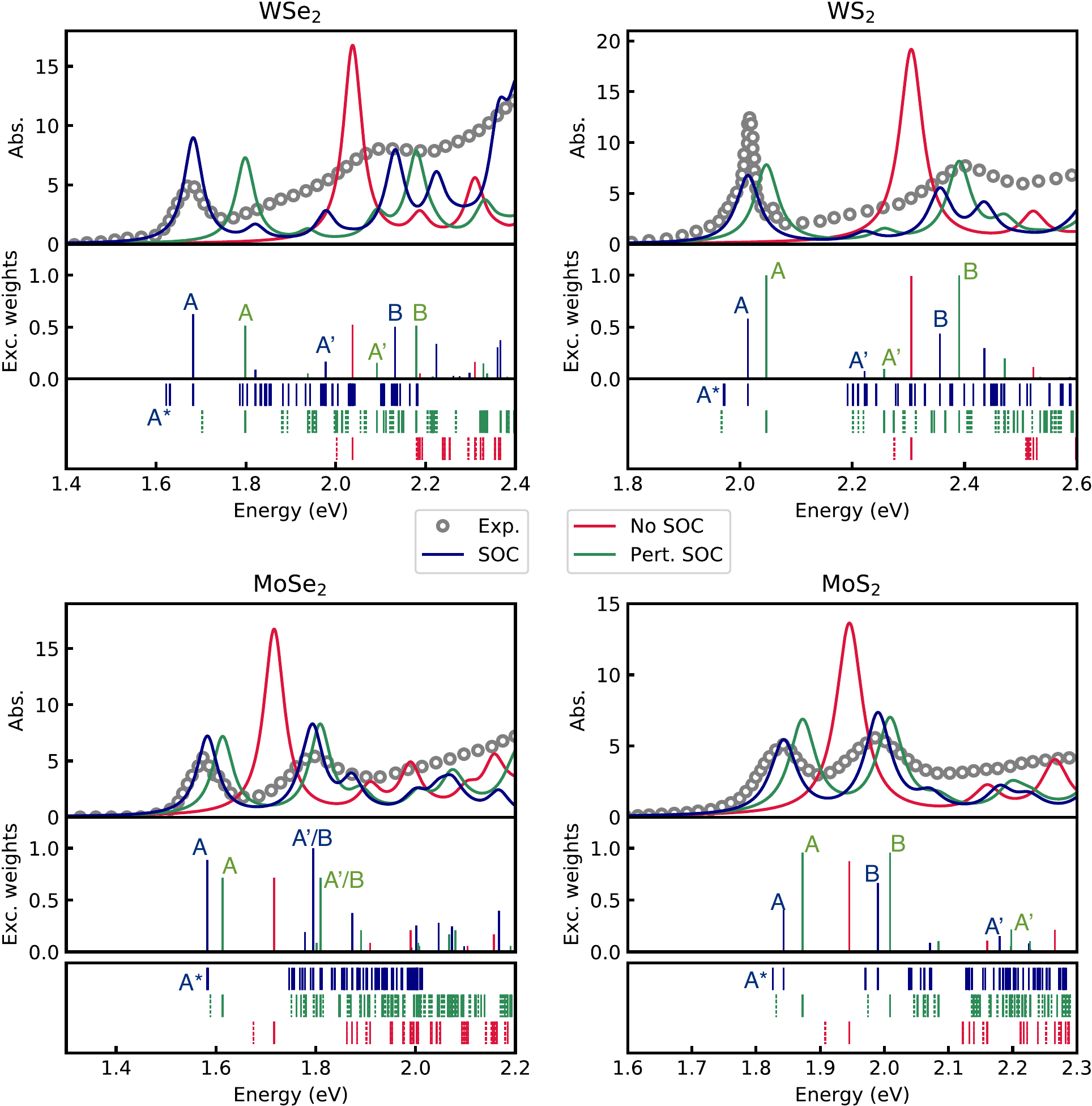}
\caption{Top panel: Bethe-Salpeter spectra of single-layer
WSe$_{2}$, WS$_{2}$, MoS$_{2}$ and MoSe$_{2}$. 
Middle panel: oscillator strenghts (normalized to 1). 
Bottom panel: each bar represents an excitation regardelss of its oscillator strength. 
No SOC, full SOC and perturbative SOC results are presented in red,  blue, and green respectively. 
All the theoretical curves have been rigidly red-shifted of $\sim$0.1-0.2 eV 
to align the energetic position of peak A to the experimental
	one. Experimental data are taken from Ref. \cite{Mak_PRL2010} for MoS$_2$, and from Ref. \cite{Kozawa2014} for the remaining systems.}
\label{fig:bse_spectra_excdos}
\end{figure*}
\begin{figure*}
\includegraphics[width=0.8\textwidth]{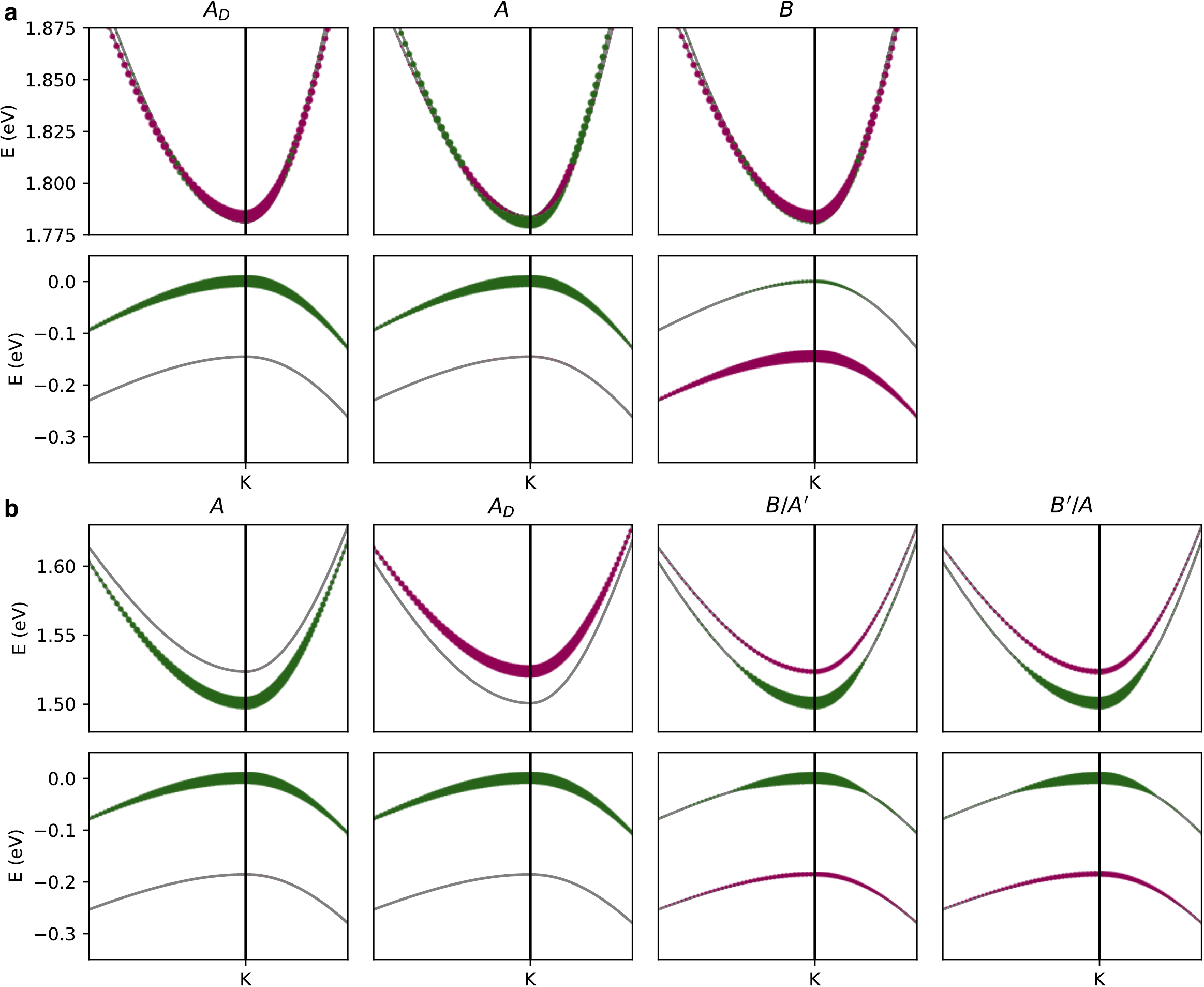}
	\caption{Zoom of the spin-splitted VBM and CBM around K showing the single particle contributions 
	to the main optical excitations: the width of the points is proportional to the contribution of 
	the state to the excitation while the color refers to the spin character. 
	Top panels:  MoS$_{2}$; bottom panels: MoSe$_{2}$
\label{fig:mo_based_bands_at_k} }
\end{figure*}
\begin{figure*}
\includegraphics[width=0.8\textwidth]{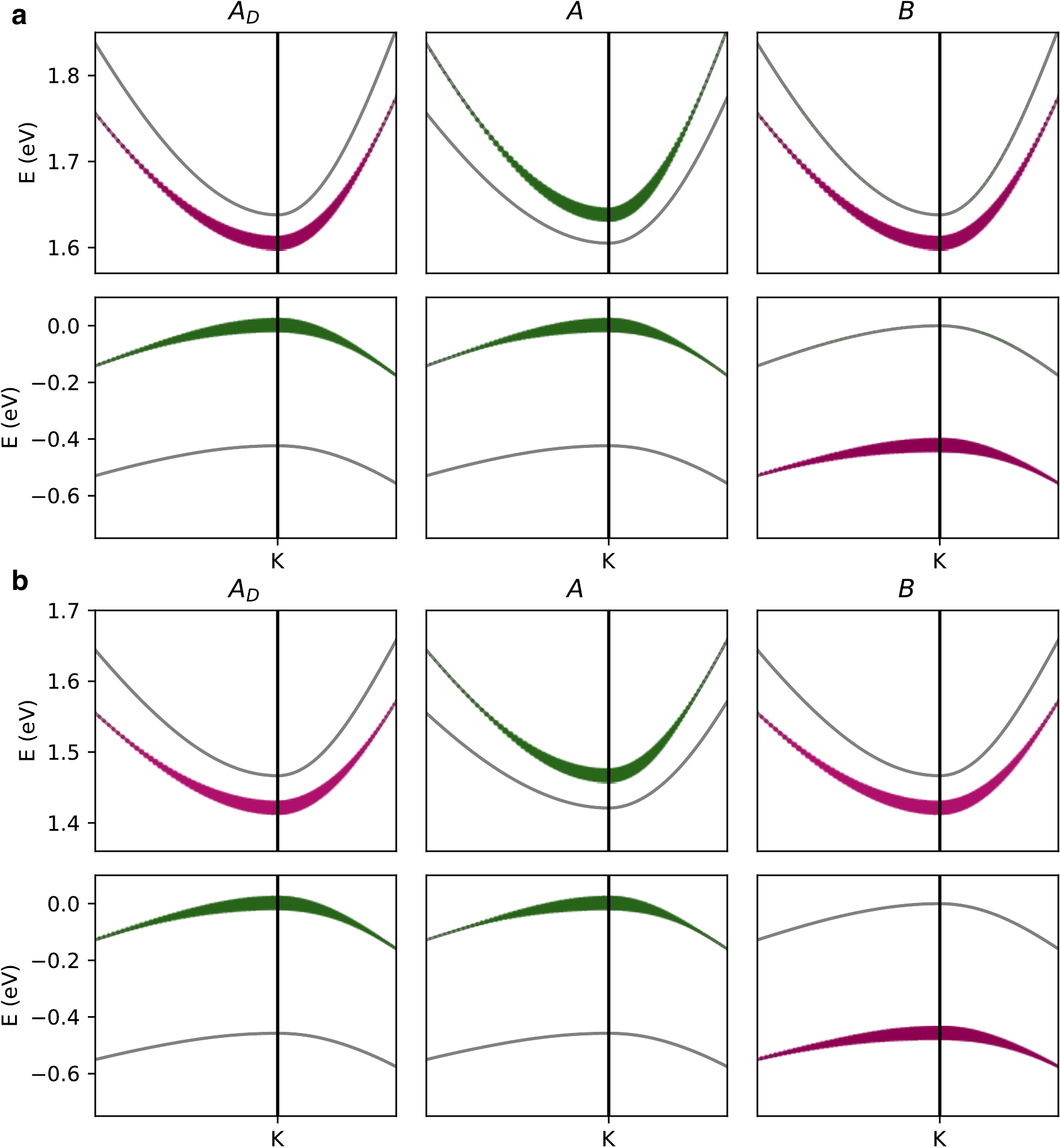}
	\caption{Zoom of the spin-splitted VBM and CBM around K showing the single particle contributions
        to the main optical excitations: the width of the points is proportional to the contribution of 
        the state to the excitation while the color refers to the spin character.
	Top panels WSe$_{2}$; bottom panels: WS$_{2}$ 
\label{fig:w_based_bands_at_k} }
\end{figure*}

In the top panels of Fig. \ref{fig:bse_spectra_excdos} we show
the GW+BSE absorbance spectra calculated without SOC and with perturbative
and non-perturbative SOC and compare them with available experimental
data \cite{Kozawa2014}. In the middle panel bars represent oscillator strengths for each excitation,
while in the bottom panel they signal the presence of an excitation, indipendently from its
oscillator strenght. 
The optical spectra are characterized by the presence of spin-splitted
strongly bound excitons and their corresponding series. The lowest
bright excitation (the so called A exciton) is mainly composed by
transitions at the six-equivalent K-points of the BZ from VBM to the
first unoccupied state with the same spin-character, which is CBM for
MoX$_{2}$ and CBM+1 for WX$_{2}$ see also Figs. \ref{fig:mo_based_bands_at_k} and 
\ref{fig:w_based_bands_at_k}. 
In Tab. \ref{tab:A_exciton} the A exciton binding energies, defined as the difference between electronic 
and optical gap, are shown for the different levels of calculation together with
the difference, between perturbative and full SOC, of the absolute position of the A peak, $\Delta E_A$, 
and of the energy of the single particle transition that mainly gives raise to it, $\Delta E_{IP}$.
In general agreement with previous
calculations \cite{Komsa12,Qiu_PRB16}, in free-standing MLs the binding energy
of exciton A are large.
By looking at the A exciton binding energy, it could be simply argued that 
perturbative SOC sistematically underbinds the A exciton
by $\sim$20 meV.  However, by taking a closer look at the absolute positions of the peak,  
the effect of applying perturbative SOC 
seems far from systematic, and thus predictable.
Indeed, when looking at $\Delta E_A$ in Tab. \ref{tab:A_exciton} and at the spectra in Fig. \ref{fig:bse_spectra_excdos},
the absolute value of the A peak position may vary by as much as 116 meV in the case of WSe$_2$.
Morever the blue shift of the A exciton in the perturbative SOC scheme, is almost entirely due 
to larger independent particle transition energies for W-based materials,
as shown by the $\Delta E_{IP}$ column of Tab. \ref{tab:A_exciton}, but not for Mo-based ones. This furthermore means
that, while for W-based materials in the perturbative and full SOC schemes 
the e-h interaction that renormalizes the independent particle (IP) transition energies is very similar, 
for Mo-based MLs perturbative SOC underestimates
the strength of e-h attraction by $\sim 20$ meV.
Finally, it is worth to mention that experimental  
binding energies depend on the substrate and are generally smaller than what 
found theoretically for free-standing layers 
\cite{Hanbicki_ssc2015,Arora_NS2015,Manca_natcomm_2017,Hill_NL2015,Liu_PRB2019,He_PRL2014,Chernikov_PRL2014,Wang_PRL2015,Chernikov2015}. Therefore the direct comparison
with theoretical data is not meaningful 
due to the presence of the substrate  
that modifies the electronic screening.

\begin{table}[!htbp]
\begin{tabular}{|l|c|c|c|c|c|}
\hline 
\multirow{2}{*}{} 
	& \multicolumn{3}{c|}{A exciton binding energy [meV]}& $\Delta E_A$ & $\Delta E_{IP}$ \tabularnewline
\cline{2-4} 
	& SOC  & Pert. SOC  & No SOC& [meV] & [meV]   \tabularnewline
\hline 
WSe$_{2}$  & 550  & 533  & 596 & 116 & 119  \tabularnewline
WS$_{2}$   & 631  & 610  & 658 & 33  & 32   \tabularnewline
MoSe$_{2}$ & 666  & 648  & 547 & 31  & 13   \tabularnewline
MoS$_{2}$  & 691  & 673  & 679 & 29  & 11   \tabularnewline
\hline 
\end{tabular}\caption{A exciton binding energy,difference between the perturbative and full
	SOC schemes of the absolute position of the A peak, $\Delta E_A$, and of the independent particle
	transition that mainly contributes to the A peak, $\Delta E_{IP}$. All energies are in (meV)}. \label{tab:A_exciton}
\end{table}


As shown in Figs. \ref{fig:mo_based_bands_at_k} and \ref{fig:w_based_bands_at_k}, 
the excitonic peaks B is mainly composed of single-particle transitions 
located at K from the spin-splitted VBM-1 to the first conduction 
band with its same spin character, namely the
CBM for W-based systems and the CBM+1 for Mo-based ones.
The IP B-A splitting, presented in Tab. \ref{tab:spectra_features}, 
arises from the DFT splitting of the conduction
and valence bands in the case of perturbative SOC, and from the spinorial GW calculation for the full SOC scheme.
Within both approaches,
the mixing of transitions reduces such IP splitting, and the perturbative SOC
reproduces quite well the full SOC B-A
splitting with the exception of WSe$_{2}$ where there is a difference
of about $\sim$70 meV. The origin of such large deviation in the case of WSe$_{2}$ can be ascribed 
to large differences already at the IP level: the IP B-A splitting difference between full and perturbative SOC
($\Delta(E_B-E_A)_{IP}$) is $\sim$80 meV 
in the case of WSe$_{2}$, while it ranges from 7 meV to 12 meV for the remaining cases. It is easy to show that 
$\Delta(E_B-E_A)_{IP}=\Delta_{GW}(VBMS)-\Delta_{GW}(CBMS)$, where $\Delta_{GW}(VBMS)$ ($\Delta_{GW}(CBMS)$ is 
the GW correction to the top (bottom) valence (conduction) band splitting. Thus, in the end, 
in WSe$_{2}$ the large difference between full and perturbative SOC B-A splittings, is due to large
GW corrections to valence and conduction band splittings which are, moreover, of opposite sign and thus sum up. On
the contrary, for WS$_2$, the GW corrections to valence and conduction band splittings have the same sign and tend to cancel
each other, finally for Mo-based systems the two contributions add up, as in WSe$_{2}$, but to a lower value, 
being one order of magnitude smaller.

In the perturbative SOC scheme exciton A and B are compelled to have the same oscillator strenght, 
which should be half of the oscillator strength of the NO SOC calculation they stem from, 
as shown in the middle panel of Fig. \ref{fig:bse_spectra_excdos}. 
On the contrary, when using a full spinorial formulation the oscillator strengths 
of the two structures may be different, yielding a better agreement with experiment 
\cite{Guo_Natphys_2019}. In particular we find that for W-based materials the oscillator 
strength of the A exciton is larger than that of the B exciton, 
while the opposite is true for Mo-based ones, although the case
of MoSe$_2$ must be handled with care, as better detailed in the following.

The A' structure is identified as the first structure following A that has similar single particle composition.
From Tab. \ref{tab:spectra_features} and Fig. \ref{fig:bse_spectra_excdos}
it can be seen that the sequence of A,B,A' peaks of the full SOC calculations is well reproduced
by the perturbative SOC approach, with the W-based materials featuring
an A' peak clearly below the B one. 
The comparison with experiments shows that the B-A splittings are well reproduced but  A'-A splittings are
in most of the cases overestimating the experimental values. 
This different behaviour may be
due to the different origin of the two splittings and to the corresponding sensitivity to the dielectric environment.
While the B-A splitting is connected to the spin-orbit splitting of the conduction and valence bands,
not affected by the changes in the electronic screening due to substrates and/or sample encapsulations, the 
A'-A splitting due the hydrogenoid series is instead stronlgy affected, just like the binding energy does.
However, it is worth noting that
the A'-A splitting may require finer k-point grids to converge its absolute value and make a meaningful comparison
with experiments, therefore we performed 
full SOC calculations on a denser 39x39x1 k-point grid for MoS$_2$, MoSe$_2$ and WSe$_2$  finding  A-A' splitting values 
of 270 meV, 231 meV, and 215 meV  respectively (further calculations on a 42x42x1 grid on WSe$_2$ show a converged
value of 215 meV). The converged values enhance the overestimation with respect to experiment suggesting that the substrate 
effect on the A-A' splitting needs further investigations, out of the scope of this paper.

\subsection{Exchange-driven intravalley mixing}
In the perturbative SOC approach, the identification of the A and B structures
is straightforward as the A and B excitons derive from the same excitation
of the underlying SOC-free BSE calculation to which different corrections are applied. 
Thus, by construction, in the perturbative SOC scheme 
the single particle transitions that contribute to the A structure are completely disjoint
from those that contribute to the B one.
However, when using a full spinorial formalism 
in the case of MoS$_2$, an exchange-driven intravalley mixing was found, 
meaning that transitions pertaining to the B exciton contributed (minimally) 
to the A exciton and viceversa \cite{Guo_Natphys_2019}. We confirm this finding  
for MoS$_2$ and found a similar mixing of contributions for 
the W-based materials. The mixing of transitions can be easily recognized looking at 
Fig. \ref{fig:mo_based_bands_at_k}, where for the B excitons we see that all bands contribute. But while
for MoS$_2$ this mixing is minimal, we find a huge effect
for the B exciton of MoSe$_{2}$.
In this case the structures that we labeled A'/B, arise from
four almost degenerate excitons which are linear combinations of
single particle transitions from the VBM to the
CBM (characterizing the A series) and from the
VBM-1 and the CBM+1 (characterizing the B series) with comparable contributions. 
This mean that in this case the exchange interaction leading to the intravalley mixing between 
VBM$\rightarrow$CBM and  (VBM-1)$\rightarrow$(CBM+1) transitions
is playing a prominent role. 
To understand this feature we must look back at the structure of the BSE Hamiltonian in the collinear magnetic case in the 
$\Delta S_z=0$ channel, see Sect. \ref{sect_BSE_blocking}. 
The significance of the exchange term is related to its magnitude with respect to the splitting of the diagonal part: 
if such splitting is small the off-diagonal exchange is able to
strongly mix the two transitions.
In the perturbative SOC calculations, where the A and B series can not mix by construction, 
the A' and B peaks are found in proximity one to the other, as also 
reported in \cite{Ugeda2014}.
The proximity in energy between these two states explains the enhanced, exchange-driven, intravalley mixing.
This finding is robust with respect to denser k-point sampling. 
A recent
upconversion experiment on a hexagonal boron nitride encapsulated sample \cite{Han_PRX_2018} is able to resolve 
two structures 150 meV and 155 meV above the A exciton, which could not be previously resolved \cite{Kikuchi_PRB_2019,Wang_2015}. 
This seems to confirm that the `accidental' proximity of the A' and B excitations is present both in our theoretical
calculations on free-standing MLs and in experiments including a dielectric environment.

\begin{table*}[!htbp]
\begin{tabular}{|c|c|c|c|c|c|c|c|c|}
\hline 
\multirow{3}{*}{}
		   & \multicolumn{5}{c|}{B-A splitting [meV]} & \multicolumn{3}{c|}{A'-A splitting [meV]} \tabularnewline
\cline{2-9}
		   & \multicolumn{2}{c|}{Pert.SOC}&\multicolumn{2}{c|}{SOC} & \multirow{2}{*}{Exp} & \multirow{2}{*}{Pert. SOC} & \multirow{2}{*}{SOC} & \multirow{2}{*}{Exp.}\tabularnewline
\cline{2-5}
		   & IP & BSE & IP & BSE &  &  &  &  \tabularnewline
\hline
	WSe$_{2}$  & 410 & 380 & 491 & 450 & 410-430\cite{Kikuchi_PRB_2019,Manca_natcomm_2017,He_PRL2014,Hanbicki_ssc2015} & 140       & 140 & 130-160 \cite{Stier_PRL2018,Manca_natcomm_2017,He_PRL2014,Wang_PRL2015,Chen_PRL2018,Liu_PRB2019} \tabularnewline 
	WS$_{2}$   & 380 & 340 & 382 & 340 & 371-395\cite{Hanbicki_ssc2015,Kikuchi_PRB_2019,Hill_NL2015} & 210       & 210 & 160 \cite{Hill_NL2015} \tabularnewline
	MoSe$_{2}$ & 210 & 200 & 221 & 200;210 & 155\cite{Han_PRX_2018} 190-220\cite{Wang_2015} & 190       & 200;210 & 155\cite{Han_PRX_2018} \tabularnewline
	MoS$_{2}$  & 150 & 140 & 160 & 150 & 124-150\cite{Robert_PRMat_2018,Kikuchi_PRB_2019,Hill_NL2015}& 210       & 230 & 175\cite{Robert_PRMat_2018}  \tabularnewline
\hline 
\end{tabular}
\caption{ A'-A and B-A splittings within full and perturbative SOC schemes. BSE and independent particle (IP) results 
are compared for the case of B-A splitting.
The B exciton in the full SOC MoSe$_{2}$ case
is not univocally defined, see discussion in the text. Experiments were either performed on a SiO$_2$/Si substrate 
\cite{Hanbicki_ssc2015,He_PRL2014,Kikuchi_PRB_2019,Wang_PRL2015,Hill_NL2015,Wang_2015},
or encapsulating the monolayer in hexagonal BN 
\cite{Manca_natcomm_2017,Liu_PRB2019,Stier_PRL2018,Chen_PRL2018,Han_PRX_2018,Robert_PRMat_2018}.
All energies are in meV. 
\label{tab:spectra_features}}
\end{table*}

\subsection{Dark-bright splitting}
Several recent experimental and theoretical works have shown that
dark (spin-forbidden and finite-momentum) excitons are present near
the first bright A exciton \cite{Malic_PhysRevMaterials2018,Molas_2dm,Zhang_natnano2017,Zhou_NN17}.
The knowledge of their energetic position is crucial in order to understand the exciton
dynamics in view of the possible use of TMD-MLs in opto-electronic devices.
Indeed at experimental level spin-forbidden dark excitons could be revealed by photoluminescence
\cite{Wang_PRL17}, ellipsometry measurements with out-of-plane light
polarization \cite{Funke} by using photocurrent spectroscopy \cite{Quereda_2dm}
or brightening with magnetic field \cite{Molas_2dm,zhang_heinz_natnano2017,LU2019}
or near-field coupling to surface plasmon polaritons \cite{Zhou_NN17}.
In particular here we focus on the lowest energy spin-forbidden dark excitons (which we
label A$^*$) in order to see
how the perturbative and full SOC schemes compare.

\begin{table*}[!htbp]
\begin{tabular}{|l|c|c|c|c|c|c|}
\hline 
\multirow{2}{*}{}
		   & \multicolumn{6}{c|}{\mbox{A-A$^*$} splitting [meV]}  \tabularnewline
\cline{2-7} 
		   & \multicolumn{5}{c|}{Theory} &  \multirow{3}{*}{Exp.}  \tabularnewline
\cline{2-6} 
		   & \multicolumn{4}{c|}{Present work} &  \multicolumn{1}{c|}{Literature} &  \tabularnewline
\cline{2-6} 
     & SOC (GW)  & SOC (scissor)  & Pert. SOC  & No SOC  & (Other)  & \\
\hline 
WSe$_{2}$  & 56  & 77  & 96  & 36  & 80 \cite{Thorsten_PhysRevB.96.201113}, 16 \cite{Echeverry2016}  & 55 \cite{Wang_PRL17}, $\simeq$ 47 \cite{Molas_2dm}, 57 \cite{zhang_heinz_natnano2017},
47 \cite{Zhou_NN17} \\
WS$_{2}$  & 42  & 72  & 80  & 30  & 80 \cite{Thorsten_PhysRevB.96.201113}, 11 \cite{Echeverry2016}  & 40 \cite{Wang_PRL17} , 47\cite{Molas_2dm} \\
MoSe$_{2}$  & -1  & 7  & 25  & 42  & 10 \cite{Thorsten_PhysRevB.96.201113}, -11 \cite{Echeverry2016}  & -1.5 \cite{LU2019}, -1.3 \cite{Natcomm_robert2020}, $\simeq$ 0 \cite{Wang_PRL17} $\simeq$ 0 \cite{Molas_2dm}
	-30 \cite{Quereda_2dm} \\
	MoS$_{2}$  & 17  & 19  & 41  & 38  & 25\cite{Thorsten_PhysRevB.96.201113}, 20 \cite{Qiu2015}, 5 \cite{Echeverry2016}  & 98 \cite{Molas_2dm}, $<$20 \cite{Funke}, 14 \cite{Natcomm_robert2020} \\
\hline 
\end{tabular}\caption{\mbox{A-A$^*$}  splitting (i.e. bright-dark splitting of exciton A).
All energies are in meV. 
\mbox{A-A$^*$} splittings from Ref. \cite{Echeverry2016} are taken from
G$_{0}$W$_{0}$-PBE + BSE calculations for Mo-based materials, while
from G$_{0}$W$_{0}$-HSE + BSE calculations for W-based ones. 
The experimental value of the \mbox{A-A$^*$} splitting for MoS$_2$ form Ref. \cite{Funke} 
was estimated from the digitalizion of Fig. 6 of the corresponding paper. Experiments were
either carried out on Si/SiO$_2$ substrate \cite{Molas_2dm,zhang_heinz_natnano2017,Quereda_2dm,Funke}, on sapphire \cite{Funke},
or on hbN encapsulated samples \cite{Natcomm_robert2020,Wang_PRL17,Zhou_NN17,LU2019}.
\label{tab:DB}}
\end{table*}

In Tab. \ref{tab:DB} the theoretical \mbox{A-A$^*$} splittings are shown and
compared with the available theoretical and experimental literature.
The No SOC scheme corresponds to the non-spin polarized calculation,
where the triplet, spin-forbidden excitons are obtained simply by switching off 
the exchange term in the BSE kernel. A full SOC BSE calculation employing a rigid shift of the conduction bands (scissor operator), opening the single-particle DFT gap to the corresponding GW value, 
was also performed. In this last case the
SOC-related splittings of the single-particle energies are the DFT ones.
Within our results we note that different theoretical schemes produce different splitting
values and that only when full-SOC is used a better agreement with
experimental data is reached. 

The case of MoS$_{2}$ is more controversial and needs further discussion. 
Magneto-photoluminescence experiments carried out on a Si/SiO$_2$ substrate
show an extremely large \mbox{A-A$^*$} splitting of $\sim$100 meV \cite{Molas_2dm}, 
whereas upon hBN ecapsulation, which is known to enhance the optical quality of the samples,  
a \mbox{A-A$^*$} splitting of 14 meV is reported \cite{Natcomm_robert2020}, also 
spectroscopic ellipsometry show a \mbox{A-A$^*$} splitting smaller than 20 meV \cite{Funke}. 
These last experimental results are more consistent with theoretical predictions,
which, within all schemes and implementations, range from 5 to 40 meV.
Indeed, the \mbox{A-A$^*$} splitting can be thought in terms of two main contributions: the first,
due to the exchange term of the BSE kernel, the second due to the CBM splitting.
The first contribution is always of positive sign and is the only present in the No SOC scheme.
The second contribution is positive for W-based systems and negative for
the Mo-based ones. In the case of the perturbative SOC and full SOC with scissor schemes
this contribution arises from the DFT CBM splitting, whereas for the full SOC calculation
it arises from the GW one.
While the exchange contribution, estimated from the No SOC scheme, is approximatively
the same for all materials, analysis of the atomic origins of the spin splitting of the conduction bands 
show that the CMB splitting is expected to be small 
for Mo-based TMDs and in particular for MoS$_{2}$ \cite{Kosmider_2013}. On this basis
the overall \mbox{A-A$^*$} splitting is expected (and theoretically predicted) smaller for Mo-based TMDs.

The scattering of the theoretical estimate of the \mbox{A-A$^*$} splitting can again be explained by the different
approach (GdW approach for \cite{Thorsten_PhysRevB.96.201113}) and/or different calculation parameters 
or numerical implementations \cite{Echeverry2016}. However, within each calculation
the picture is consistent with a more or less material-independent exchange contribution plus material specific CBM,CBM+1 splittings 
(see Tab. \ref{tab:band_splittings} and Ref. \cite{Echeverry2016}).

\subsection{Excitonic spin polarization}
\begin{figure*}[t]
\includegraphics[width=0.8\textwidth]{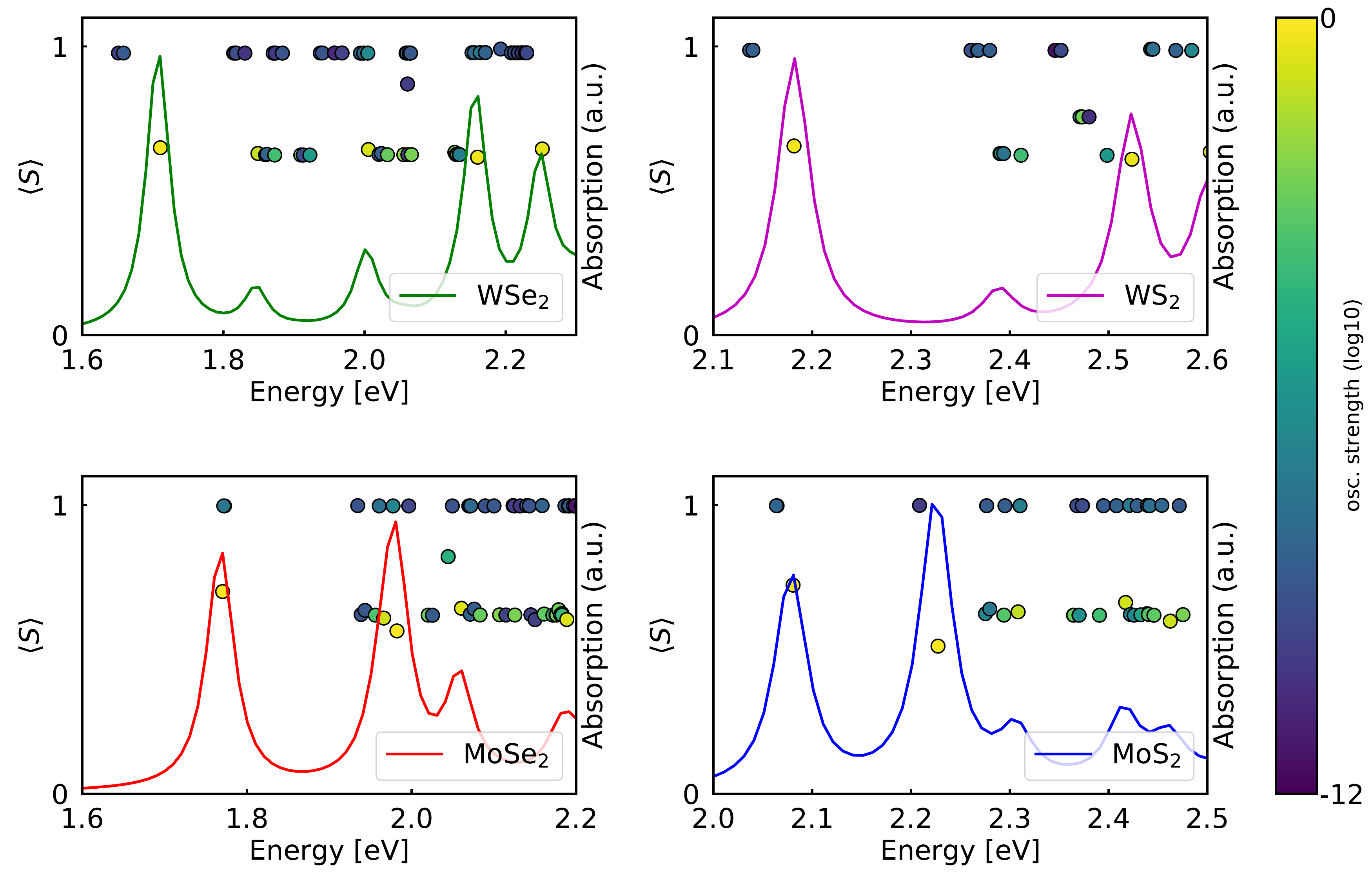}
\caption{Dots: excitonic total spin expectation value. The dot color refers
to the oscillator strength of that specific excitation (normalized
to one). In full lines, as a reference for the eye, the absorption
spectra. Top panel: W-based materials: WSe$_{2}$ left, WS$_{2}$
right. Bottom panel: Mo-based materials: MoSe$_{2}$ left, MoS$_{2}$
right. \label{fig:excitonic-total-spin} }
\end{figure*}

In a full spinorial treatment it is possible to determine the exciton's
spin expectation value, namely $\langle S^2\rangle$, given by Eq. \ref{eq:exc_spin_10},
and its projections $N_{SM}^{\lambda}$ along the singlet state and
along the three components of the triplet state, given by Eq. \ref{eq:exc_spin_9}.
It is worth noting that such analysis is not possible in a perturbative
SOC approach where only the excitation energies are corrected and
the single particle wavefunctions are kept fixed to the non magnetic
case. In Fig. \ref{fig:excitonic-total-spin} the dots represent $\langle S\rangle$
for each excitation, the color of the dot is determined by its oscillator
strength, while the absorption spectrum is plotted as
a guide for the eye in a full line. There are two main classes of excitations those
whose $\langle S\rangle\sim1$ and those whose $\langle S\rangle\sim0.6$.
All the $\langle S\rangle\sim1$ excitons are dark: as they are mainly
composed of linear combinations of the $|S=1,M=\pm1\rangle$ states
and are thus spin-forbidden excitations. The situation for the $\langle S\rangle\sim0.6$
is less homogeneous, some excitations are bright and build up the
optical spectrum, while others are dark. In this case the origin of
the low oscillator strength can not be addressed to spin conservation
but to dipole symmetry rules. The fact that $\langle S\rangle\sim0.6$,
significatively lower than $1$ is an evidence of a strong component
along the singlet state. 

Indeed in Tab. \ref{tab:Excitonic-spin-analysis.} we report the decomposition
of the main structures identified in the absorption spectrum, along
with that of the dark A excitation. The A and B excitons are linear
combinations of the $|S=0,M=0\rangle$ and $|S=1,M=0\rangle$ states,
whereas the dark A peak is made mainly of $|S=1,M=\pm1\rangle$ contributions. 

First of all it is possible to notice that $\langle S_{z}\rangle=0$ for all the excitations, including the dark ones 
related to the magnon channel, this is a consequence of the symmetry between the K and K' valley which equally contribute.
Secondly, in principle in a full spinorial, non collinear, approach all the $|S,M\rangle$
components would be allowed to mix, while, as discussed in Sect. \ref{sec:BSE_blocking} and shown in Ref. \cite{rodl_prb2008},
in a collinear but spin polarized case, only the singlet and the $M=0$ component
of the triplet
are. The spin analysis of the low energy excitations
shows that here we are somewhat close to this case: the spinor states contributing 
to the low energy excitations are strongly spin polarized and thus the mixing of the singlet and $M=\pm1$ triplet components
is very small.  
Moreover SOC removes the degeneracy between the single-particle 
contributions pertaining to the A and B excitons, at the same time, this SOC splitting is much larger
than the exchange contribution which couples such transitions. As a consequence, the the $|S=0,M=0\rangle$ and
$|S=1,M=0\rangle$ states are strongly mixed and 
the triplet or singlet character of the excitations is destroyed
(see also App.~\ref{App:SOC_and_exchange}). Finally, the fact that for the bright excitations 
the weights of the $|S=0,M=0\rangle$ and $|S=1,M=0\rangle$
components are not exactly equal, as shown in Tab. \ref{tab:Excitonic-spin-analysis.}, 
is an effect of the small but finite exchange term. As shown in \cite{Guo_Natphys_2019} for MoS$_2$, and found here also 
for the remaining cases, exchange is responsible for the intravalley mixing of the A and B single particle contributions.

\begin{table*}[!htbp]
\begin{tabular}{|c|c|c|c|c|c|c|}
\hline 
	& $\langle S^2\rangle$ & $\langle S_{z}\rangle$ & $|S=0,M=0\rangle$ & $|S=1,M=-1\rangle$ & $|S=1,M=0\rangle$ & $|S=1,M=1\rangle$\tabularnewline
\hline 
\hline 
\multicolumn{7}{|c|}{WSe$_{2}$}\tabularnewline
\hline 
	A dark&	1.899&	0&	0.034&	0.466&	0.034&	0.466\tabularnewline
\hline 
	A&	0.823&	0&	0.464&	0.007&	0.522&	0.007\tabularnewline
\hline 
	A'&	0.776	&0&	0.487&	0.007	&0.499&	0.007\tabularnewline
\hline 
	B	&0.748	&0	&0.502&	0.037&	0.425&	0.037\tabularnewline
\hline 
\multicolumn{7}{|c|}{WS$_{2}$}\tabularnewline
\hline 
	A dark&	1.946&	0&	0.019&	0.481&	0.02&	0.481\tabularnewline
	A&	0.838&	0	&0.458	&0.006	&0.531&	0.006\tabularnewline
	A'	&0.776	&0	&0.488&	0.006&	0.501&	0.006\tabularnewline
	B	&0.730	&0	&0.51	&0.022	&0.446	&0.022\tabularnewline
\hline 
\multicolumn{7}{|c|}{MoSe$_{2}$}\tabularnewline
\hline 
	A&	0.951	&0	&0.404&	0.001&	0.594&	0.001\tabularnewline
	A dark	&1.988&	0	&0.004	&0.496	&0.004	&0.496\tabularnewline
	A' (B)&	0.730	&0	&0.51&	0.003&	0.484&	0.003\tabularnewline
	A''	&0.635	&0	&0.559	&0.003	&0.435	&0.003\tabularnewline
	B'	&0.718	&0	&0.516	&0.005	&0.474	&0.005\tabularnewline
\hline 
\multicolumn{7}{|c|}{MoS$_{2}$}\tabularnewline
\hline 
	A dark	&1.991&	0&	0.003	&0.497	&0.003	&0.497\tabularnewline
	A	&1.011&	0&	0.377&	0.001&	0.621&	0.001\tabularnewline
	B dark	&1.997	&0	&0.001	&0.499	&0.001	&0.499\tabularnewline
	B	&0.535	&0	&0.614	&0.003	&0.38	&0.003\tabularnewline
	A'	&0.778	&0	&0.486	&0.001	&0.512	&0.001\tabularnewline
	B'	&0.705	&0	&0.522	&0.003	&0.471	&0.003\tabularnewline
\hline 
\end{tabular}

\caption{Excitonic spin analysis: $\langle S\rangle$, and $\langle S_z\rangle$   
expectation values, 
and the projections of the excitonic wave functions along the singlet state and
along the three components of the triplet state.	
 \label{tab:Excitonic-spin-analysis.} }
\end{table*}

\section{conclusions}\label{sec:CONCLUSIONS}
In this paper, a detailed derivation of the GW and BSE equations  by including
the full spinorial nature of the wavefunctions, is illustrated.
This formulation allows to obtain the collinear, spin and non-spin polarized, and the non collinear cases in a natural way.
The spin-orbit interaction can then be included in a non perturbative way from the ground-state up to the excited state simulations.
Beyond the formal derivation of all the main equations,  we carry out a systematic analysis of electronic and optical properties
of most representative group VI TMD monolayers, comparing at the same level of numerical implementation, results without SOC and with SOC at perturbative and non perturbative level.
While in most of the observables considerered the perturbative and non perturbative approach for SOC give very similar results,
the dark-bright splittings are generally improved when the SOC is included in a non pertubative way. The exchange-driven intravalley mixing, absent by construction within the perturbative approach, is found to hugely impact the 
nature of the B exciton for the case of MoSe$_2$ that is found to be strongly mixed with the A' exciton.
Furthermore the spin character of all the excitons in the IR-Vis region for the four TMDs is analyzed, which is impossible in a perturbative SOC approach,
and not yet available, has been obstained and discussed

\begin{acknowledgments}
MP acknowledges funding from INFN20-TIME2QUEST project.
MM aknwoledges that part of the computing resources 
and the related technical support used for
this work have been provided by CRESCO/ENEAGRID High Performance
Computing infrastructure and its staff \cite{cresco}. 
CRESCO/ENEAGRID High
Performance Computing infrastructure is funded by ENEA, the Italian
National Agency for New Technologies, Energy and Sustainable Economic
Development and by Italian and European research programmes, see
http://www.cresco.enea.it/english for information". 

\end{acknowledgments}

\appendix
\section{Spin dependent Exchange--Correlation in a spinorial basis}~\label{App:DFT_and_SOC}

The density matrix and the xc-potential can be written in terms of the density ($n$) and the magnetization ($\mathbf{m}$) as:
\begin{subequations}
\label{spin-magn-relation}
\begin{gather}
\uu{\gr}\(\xx\)=n\(\xx\)\uu{\gs}_0+\mathbf{m}\(\xx\)\cdot\uu{\pmb{\gs}}, \\
\uu{v}^{xc}\(\xx\)=\phi^{xc}\(\xx\)\uu{\gs}_0+\mathbf{B}^{xc}\(\xx\)\cdot \uu{\pmb{\gs}}, 
\end{gather}
\end{subequations}
where we also introduced the exchange--correlation magnetic field, $\vec{B}^{xc}$ and density-potential $\phi^{xc}$.

Although known xc functionals are (local) functions of the modulus of $\mathbf{m}$ alone, spin dependent KS equations can be solved also in the LDA for a non-collinear system. This is obtained by calculating $\uu{v}^{KS}$,
which depends in the LDA only from the magnitude of the local magnetization,
by rotating the magnetization vector in each point in space into the local frame of spin-quantization
and evaluating the potential $v_{xc}(n(r),m(r))$ and than rotating back to the global reference frame.

\section{The Hedin's equations in spin, space and time representation}~\label{App:Hedin_textbook}
The main ingredient of the approach is the  electronic Green's Function\,(GF):
\begin{align}
 G\(1,2\)=-i\Langle\mathcal{T}\wl \hat \Psi\(1\)\hat\Psi^\dagger\(2\) \wr\Rangle,
  \label{eq:3.2}
\end{align}
where $\langle\ldots\rangle$ is the trace evaluated with the exact density matrix, $1=(\xx_1,s_1,t_1)$
includes space, spin and time, and operators are in the Heisenberg representation. 

From \e{eq:dft.1} it follows that the non--interacting is the KS one:
\eq{
 G^{\(0\)}\(1,2\) \equiv G^{KS}\(1,2\).
\label{eq:dft.3}
}

The Hedin's equations can be derived by using the functional derivatives Schwinger
approach~\cite{Strinati1988} where $H$ is perturbed with a spin collinear 
\emph{time-dependent auxiliary field} $\eta\(1\)$ 
\begin{align}
  \hat{H}_{\eta}\(t_1\)=\hat{H} + \int\!\!d\xx_1\, \eta\(1\) \hat\Psi^\dagger(1) \hat\Psi(1).
\label{eq:4.1}
\end{align}
We can safely use a collinear perturbation because, as explained in Ref.\onlinecite{Aryasetiawan2008}, this is consistent with the fact that the Coloumb
interaction is spin independent. More elaborated auxiliary fields must be introduced in the case of non--collinear electron--electron mediated interactions.

It can be easily proved~\cite{Aryasetiawan2008,Strinati1988} that $G$ solves a set of self--consistent, integro--differential  equations. The Hedin's equations, 
The first equation is the usual Dyson equation 
\begin{align}
G\(1,2\) =G^{\(0\)}\(1,2\) + G^{\(0\)}\(1,3\)\Sigma^{Hxc}\(3,4\) G\(4,2\),
\label{eq:4.4}
\end{align}
with repeated subscripts summed and repeated arguments integrated, if not explicitly written. 

In \e{eq:4.4} $\Sigma^{Hxc}$ is composed of two terms:
\begin{align}
 \Sigma^{Hxc}\(3,4\)=\Sigma\(3,4\)+v^{H}\(3\)\gd\(3,4\),
\label{eq:4.4a}
\end{align}
with $v^{H}$ the Hartree potential,
\begin{align}
 v^{H}\(3\)=v^{H}\(\xx_3\)=-iv\(\xx_3-\xx_5\)G\(5,5^{+}\),
 \label{eq:4.4b}
\end{align}
and $\Sigma$ the exchange and carrelation self-energy also known as the Mass operator. $\Sigma$ can be rewritten in terms of an irreducible vertex function $\wt{\Gamma}$:
\begin{align}
\label{eq:4.5}
\Sigma\(1,2\)=  -i G\(1,3\) \wt{\Gamma}\(3,2;4\) W\(4,1\),
\end{align}
with,
\begin{multline}
\label{eq:4.6}
\wt{\Gamma}\(1,2;3\)=-\frac{\gd G^{-1}\(1,2\)}{\gd \eta\(3\)}
                   =\gd\(1,3\)\gd\(2,3\)+\\\frac{\gd \Sigma\(1,2\)}{\gd G\(4,5\)} G\(4,6\)\wt{\Gamma}\(6,7;3\) G\(7,5\).
\end{multline}
$W\(4,1\)$ is the electronic screened interaction
\begin{align}
\label{eq:4.7}
W\(4,1\)=v\(4,1\)+v\(4,5\) \wt{\chi}\(5,6\) W\(6,1\),
\end{align}
that is written in terms of $\wt{\chi}$, the irreducible electronic response function:
\begin{align}
\label{eq:4.8}
\wt{\chi}\(5,6\)= G\(5,7\) \wt{\Gamma}\(7,8;6\) G\(8,5\).
\end{align}
Eqs.(\ref{eq:4.5}--\ref{eq:4.8}) represent the spin Hedin's equations and completely solve the many--body problem.  

Starting from the equation for the response function and the vertex
a Dyson like equation for a response function appear. However, due to the structure of the variables, such equation cannot be
directly cast in terms of the two point response function (not even with the approximation defined in Eq.~\eqref{eq:bse.2}).
It must be cast in terms of the four point function $L(1,3;2,4)$. Defining $K=\partial\Sigma/\partial G$ and inserting
Eq.\eqref{eq:4.6} into Eq.\eqref{eq:4.8} one obtaines
\begin{multline}
L(13,24)=L_0(13,24)+\\+L_0(13,1'3') K(1'3',2'4') L(2'4',24)
\end{multline}
where $L_0=GG$ and $L(11,22)=\chi(1,2)$. 

\section{Rotation in the spinorial basis of the different components of Hedin's equations}~\label{app:rotation}
The two maps needed to rotate Hedin's equations in the spinorial basis are defined in \e{eq:5.0}.
The goal of the different sections of this appendix is to demonstrate how the two maps follow from the manipulation of Hedin's equations.

\subsection{The Dyson equation}~\label{app:dyson}
The transformation of the self--energy operator follows easly by taking \e{eq:4.4} and expanding both $G$ and $G^{\(0\)}$ using $M_2$. It follows that
\begin{multline}
\Sigma^{Hxc}_{\II_1 \II_2}\(t_1,t_2\) = \phi^*_{\II_1 s_1}\(\xx_1\) \Sigma^{Hxc}\(1,2\) \phi_{\II_2 s_2}\(\xx_2\)\\
=\Sigma_{\II_1 \II_2}\(t_1,t_2\)+v^{H}_{\II_1 \II_2}\(t_1\)\gd\(t_1-t_2\).
\label{eq:app_dyson.1}
\end{multline}
$v^{H}_{\II_3 \II_2}$ is defined in \e{eq:app_h.2}, while $\Sigma_{\II_1,\II_2}$ is defined in \e{eq:map_Sigma_explicit}.

\subsection{The Hartree Potential}~\label{app:Hartree}
From \e{eq:app_dyson.1} it follows that
\begin{align}
v^{H}_{\II_3 \II_4}\(t_3\) = \phi^*_{\II_3 s_3}\(\xx_3\) v^{H}\(3\) \phi_{\II_4 s_4}\(\xx_3\).
\label{eq:app_h.1}
\end{align}
By using \e{eq:4.4b} we see that 
\begin{align}
v^{H}_{\II_3 \II_4}\(t_3\) = -i V_{  \substack{ \II_3 \II_4 \\ \II_5 \II_{5'}} }  G_{\II_5 \II_{5'}},
\label{eq:app_h.2}
\end{align}
with
\begin{multline}
V_{ \substack{ \II_1 \II_2 \\ \II_3 \II_4} }=
\phi^*_{\II_1 s_1}\(\xx_1\) \phi_{\II_1 s_1}\(\xx_1\) \\ \times v\(\xx_1-\xx_3\) \phi^*_{\II_3 s_3}\(\xx_3\) \phi_{\II_4 s_3}\(\xx_3\).
\label{eq:app_h.3}
\end{multline}
\e{eq:app_h.3} is the proof of \e{eq:map_v}.

\subsection{The Vertex Function}~\label{app:vertex}
In order to rotate the vertex function we observe that, from \e{eq:map_G_explicit} it follows that $G^{-1}_{\II_1 \II_2}$ rotates like $\Sigma_{\II_1 \II_2}$. This implies that
we can rewrite 
\begin{multline}
\wt{\Gamma}\(3,2;4\)= -\phi_{\II_3 s_3}\(\xx_3\) \phi^*_{\II_2 s_2}\(\xx_2\)\\\times
 \frac{\gd G^{-1}_{\II_3 \II_2}\(t_3,t_2\)}{\gd \eta_{\II_4 \II_5}\(t_5\)}
 \frac{\gd \eta_{\II_4 \II_5}\(t_5\) }{\gd \eta\(4\)},
\label{eq:app_v.1}
\end{multline}
where we have introduced
\begin{align}
\eta_{\II_4 \II_5}\(t_5\) = \phi_{\II_4 s_5}\(\xx_5\) \eta\(5\) \phi^*_{\II_5 s_5}\(\xx_5\),
\label{eq:app_v.2}
\end{align}
and
\begin{align}
\wt{\Gamma}_{  \substack{ \II_3 \II_2 \\ \II_4 \II_5} }\(t_3,t_2;t_4\)\equiv \frac{\gd G^{-1}_{\II_3 \II_2}\(t_3,t_2\)}{\gd \eta_{\II_4 \II_5}\(t_4\)}.
\label{eq:app_v.3}
\end{align}
From \e{eq:app_v.2} the functional derivative appearing on the r.h.s. of \e{eq:app_v.1} can be easily calculated to give
\begin{multline}
\wt{\Gamma}\(3,2;4\)= - 
 \phi_{\II_3 s_3}\(\xx_3\) \phi^*_{\II_2 s_2}\(\xx_2\)\\\times
 \phi_{\II_4 s_4}\(\xx_4\) \phi^*_{\II_5 s_4}\(\xx_4\)
 \frac{\gd G^{-1}_{\II_3 \II_2}\(t_3,t_2\)}{\gd \eta_{\II_4 \II_5}\(t_4\)}.
\label{eq:app_v.5}
\end{multline}
\e{eq:app_h.3} is the proof of \e{eq:map_gamma}.

\subsection{The Response Function}~\label{app:X}
The rotation of  the response function follows from \e{eq:4.8}, after using \e{eq:5.0}  and \e{eq:app_v.5}. It follows that
\begin{multline}
\wt{\chi}\(1,2\) = 
 \phi_{\II_1 s_1}\(\xx_1\) \phi^*_{\II'_1 s_1}\(\xx_1\)  \\ \times 
 \wt{L}_{  \substack{ \II_1 \II'_1 \\ \II_2 \II'_2} } \(t_1,t_2\)  \phi_{\II_2 s_2}\(\xx_2\) 
 \phi^*_{\II'_2 s_2}\(\xx_2\),
\label{eq:app_X.1}
\end{multline}
which, using \e{eq:4.7} demonstrates \e{eq:map_chi_explicit}.
The response function is a particlar case.
Indeed ${\wt{L}_{  \substack{ \II_1 \II_{1'} \\ \II_2 \II_{2'}} } \(t_1,t_2\)\neq \wt{\chi}_{  \substack{ \II_1 \II_{1'} \\ \II_2 \II_{2'}} } \(t_1,t_2\)}$ since
$\wt{\chi}_{  \substack{ \II_1 \II_{1'} \\ \II_2 \II_{2'}} } $, which results from $M_4 : \wt{\chi}$ can only be used to reconstruct $\wt{\chi}(1,2)$ via the inversion of $M_4$ while
$\wt{L}_{  \substack{ \II_1 \II_{1'} \\ \II_2 \II_{2'}} } \(t_1,t_2\)$ contain enough information to re-construct both $\wt{\chi}(1,2)$ and $\wt{L}(1,2)$.
However, when the Dyson equation for the response function is written in the spinorial representation (or more in general in a wave--function basis--set)
the contraction is not present anymore and the matrix elements of $L$ appears.

\subsection{The exchange--correlation self--energy operator}~\label{app:mass}
If now we use \e{eq:map_G_explicit} and \e{eq:app_v.5} to expand $G$ and $\wt{\Gamma}$ in \e{eq:4.5} for $\Sigma$, we get
\begin{multline}
\Sigma_{\II_1,\II_2}\(t_1,t_2\) = -i  G_{\II_{1'} \II_3}\(t_1,t_3\) \wt{\Gamma}_{  \substack{ \II_3 \II_2 \\ \II_4 \II_{4'}} }\(t_3,t_2;t_4\)\\\times
\[ \phi^*_{\II_1 s_1}\(\xx_1\) \phi_{\II_{1'} s_1}\(\xx_1\) W\(4,1\) 
 \phi^*_{\II_{4} s_4}\(\xx_4\) \phi_{\II_{4} s_4}\(\xx_4\) \].
\label{eq:app_m.1}
\end{multline}
which implies \e{eq:map_Sigma_explicit}.

\section{Computational Details}~\label{App:compdetails}
The Density Functional Theory (DFT) 
simulations have been performed using the plane-wave Quantum-Espresso code \cite{Giannozzi2009}. A Perdew-Burke-Ernzerhof (PBE) exchange-correlation functional~\cite{PBE}
and  
optimized norm-conserving \cite{hamann2013optimized} pseudopotentials have been used.
A $16\times16\times1$ Monkhorst-Pack grid \cite{monkorst-pack} of $k$-points are used to sample the Brillouin zone for structural optimization runs. A kinetic energy cutoff of 140 Ry has been used. Structure relaxation is assumed at convergence when the maximum component of the residual forces on the ions is smaller than 10$^{-5}$ Ry/Bohr.  
The lattice parameters of the four systems are in very good agreement with existing literature being a=3.12 \AA (for MS$_2$), 3.25 \AA (for MSe$_2$), 3.12 \AA (for W$_2$), and 3.25 \AA (for WSe$_2$).
Once the optimized atomic structures have been obtained, self and non-self consistent DFT calculations have been  performed to obtain Kohn-Sham (KS) eigenvalues and eigenfuctions to be used in the many-body simulations done by using the many-body code YAMBO \cite{AndreaMarini2009,Sangalli_2019}.
Specifically, we calculated the quasi-particle (QP) energies by using the $GW$ perturbative one-shot method and the optical excitation energies and the optical spectra  by solving the Bethe--Salpeter Equation (BSE) \cite{SMH1982,Strinati1988,HS1974,HS1980,Onida2002}.
For $GW$ simulations a plasmon--pole approximation for the inverse dielectric matrix has been applied energy\cite{godby1989metal}, kinetic energy cutoff of 10 Ry (60 Ry) are used for the  correlation, $\Sigma_{c}$, (exchange, $\Sigma_{x}$) part of the self-energy and the sum over the unoccupied states for $\Sigma_{c}$ and the dielectric matrix is performed up to $\sim$ 30 eV above the VBM; in order to speed up the convergence with respect to empty states we adopted the technique described in Ref.~\citenum{bruneval2008accurate}.
The Bethe-Salpeter equation to obtain the optical spectrum and exciton spatial localization has been solved within the Tamm-Dancoff \cite{Dancoff,Onida2002} approximation (which is generally valid for bulk compounds to describe neutral excitations well below the plasma frequency of the material). 2 occupied and 2 unoccupied states have been used to build up the excitonic Hamiltonian.  
For both the $GW$ and $BSE$ simulations we used a k-grid of 33x33x1 which is enough to provide a good convergence in the position of the A-B excitons, which are the main focus of this manuscript.

\section{Perturbative SOC and exchange}~\label{App:SOC_and_exchange}

It is important to notice that there are two different cases to be considered for spin-conserving excitations (i.e. $\Delta S_z=0$)
\begin{itemize}
	\item[(a)] The singlet--triplet splitting, i.e. the exchange splitting $\Delta E_\lambda^{T-S}$ due to $H^{exch}$, is bigger that the SOC correction.
	\item[(b)] SOC is, on one side, small enough to be considered a perturbation and, on the other side, significantly bigger than the $\Delta E_\lambda^{T-S}$ splitting.
\end{itemize}

In case (a) we expect the singlet and triplet excitation to remain far in energy.
The SOC splitting just corrects such energies. Thus it makes sense to define the pSOC as an energy shift in the S (and T) channel. This is the scheme reported in the literature and in the main text.

In case (b) instead 
we expect the singlet and triplet structure to be destroyed.
Neither use the singlet hamiltonian, 
with ${H^{exch,S}=2V}$, nor the triplet one,
with ${H^{exch,T}=0}$, should be used, but an average hamiltonian with
${H^{exch}=V}$. The resulting eigenvectors $\tilde{A}^{\lambda}_{cv\kk}$ define
\begin{eqnarray}
\Psi^{\lambda,s}_{ss'}(\xx_c,\xx_v)=\delta_{s,s'}\sum_{cv\kk}\tilde{A}^{\lambda}_{cv\kk}\phi^{0,*}_{c\kk s}(\xx_c)\phi^0_{v\kk s}(\xx_v),
\end{eqnarray}
where now $s$ is considered as an approximate quantum number.
The two associated transitions would then be
\begin{eqnarray}
\Delta\omega_{\lambda,s}^{0} \approx \sum_{cv\kk} |\tilde{A}^{\lambda}_{cv\kk}|^2 (\Delta \epsilon^{SOC}_{c\kk s}-\Delta \epsilon^{SOC}_{v\kk s}) 
\label{eq:pert_soc_singlets}
\end{eqnarray}
The final result is that each peak is splitted in two peaks with equal intensity both belonging to the
$\Delta S_z=0$ channel, one for $\up-\up$ transitions and the other for $\dn-\dn$ transitions. Accordingly $S$ is not anymore a good quantum number for the exciton
and one would expect $\langle \hat{S}\rangle \neq {0, 1}$.

\section{Mapping and degenerate states}~\label{App:mapping_degenerate}
	
In case of degeneracy we define the subgroups of degenerate spaces as $G^\kk_{m s}$ and $G^\kk_{n}$. We observe that 
the states  without SOC, $|m\kk s\rangle$, are more degenerate than the states with SOC, $|n\kk \rangle$, such that to each $G^\kk_{m s}$ may correspond multiple $G^\kk_{n}$. Then we define, for each $n\kk$
\begin{eqnarray}
|\Delta^{\kk}_{n,G^{\kk}_{ms}}|^2&=&\sum_{ms\in G^{\kk}_{ms}} |\Delta^{\kk}_{n,ms}|^2 \\
D^\kk_n&=&\max_{G^{\kk}_{ms}} |\Delta^{\kk}_{n,G^{\kk}_{ms}}|^2.
\end{eqnarray}
The mapping function is then defined as
\begin{equation}
f_\kk(ms)=n \quad\text{if}\quad  |\Delta^{\kk}_{n,G^\kk_{ms}}|^2= D^\kk_{n},
\end{equation}
where the mapping is done recursively, doing a loop over the $n$ index, and we assign $n$ randomly to one of the states ${ms\in G^\kk_{ms}}$ which has not any other $n$ assigned. Moreover, if $n \in G^\kk_n$, the procedure picks up the ${n\in G^\kk_n}$ with maximum projection in ${G^\kk_n}$, which is again random. Since we are dealing with degenerate energies, these choices do not affect the final result.
At the end, there is one $|\orar{n\kk}\rangle$  state assigned to each $|\orar{m\kk s}\rangle$ state.


%
%
\end{document}